\documentclass[a4paper,12pt]{article}

\pdfoutput = 1

\usepackage{graphicx}
\usepackage[numbers]{natbib}
\usepackage{amssymb}

\usepackage{amsthm}
\usepackage{amsmath}
\usepackage[latin1]{inputenc}
\usepackage{epsfig}
\usepackage{epstopdf}
\usepackage{fancybox}
\usepackage{float}
\usepackage{pifont}
\usepackage{fancyhdr}
\usepackage{url}
\usepackage{color}
\usepackage{placeins}  
\usepackage{setspace}  
\usepackage{lastpage}
\usepackage{array}
\usepackage[figure,table]{totalcount}

\def \RR {\mathbb{R}}
\def \EE {\mathbb{E}}

\def\bx{\boldsymbol{x}}
\def\bell{\boldsymbol{\ell}}
\def\by{\boldsymbol{y}}

\def\bh{\boldsymbol{h}}

\def\br{\boldsymbol{r}}

\def\I{{\bf I}}

\def\R{{\bf R}}

\def\U{{\bf U}}
\def\S{{\bf S}}

\DeclareMathOperator*{\argmin}{arg\,min}

\title{Improvement of code behaviour in a design of experiments by metamodeling}

\author{Fran\c cois Bachoc\thanks{Corresponding author. E-mail: francois.bachoc@math.univ-toulouse.fr Address: Institut de Math\'ematiques de Toulouse, Universit\'e Paul Sabatier, 118 route de Narbonne, 31062 TOULOUSE Cedex 9, France.
Phone: 0033 5 61 55 69 16}\hspace{0.2cm}\thanks{The author conducted a part of the research related to this manuscript when he was affiliated first to CEA-Saclay, DEN, DM2S, STMF, LGLS, F-91191 Gif-Sur-Yvette, France and then to the University of Vienna.} \\
{ \small Institut de Math\'ematiques de Toulouse} \\
Karim Ammar \\
{ \small CEA-Saclay, DEN, DM2S, SERMA, LPEC, F-91191 Gif-Sur-Yvette, France} \\
Jean-Marc Martinez \\
{ \small CEA-Saclay, DEN, DM2S, STMF, LGLS, F-91191 Gif-Sur-Yvette, France} 
}

\date{}

\begin{document}

\maketitle

\vspace{0.5cm}

{ \large \textbf{Total number of pages: \pageref{LastPage}}}

\vspace{0.5cm}

{ \large \textbf{Total number of tables: \totaltables  }}

\vspace{0.5cm}

{ \large \textbf{Total number of figures: \totalfigures } }


\newpage

\doublespacing

\begin{center}
\Large{Improvement of code behaviour in a design of experiments by metamodeling}
\end{center}

\vspace{0.5cm}

\begin{center}
Fran\c cois Bachoc (IMT), Karim Ammar (CEA), Jean-Marc Martinez (CEA) 
\end{center}

\vspace{0.5cm}

\begin{abstract}

It is now common practice in nuclear engineering to base extensive studies on numerical computer models. These studies require to run computer codes in potentially thousands of numerical configurations and without expert individual controls on the computational and physical aspects of each simulations.
In this paper, we compare different statistical metamodeling techniques and show how metamodels can help to improve the global behaviour of codes in these extensive studies. We consider the metamodeling of the Germinal thermalmechanical code by Kriging, kernel regression and neural networks. Kriging provides the most accurate predictions while neural networks yield the fastest metamodel functions. All three metamodels can conveniently detect strong computation failures. It is however significantly more challenging to detect code instabilities, that is groups of computations that are all valid, but numerically inconsistent with one another. For code instability detection, we find that Kriging provides the most useful tools.

\end{abstract}

\vspace{1cm}


\newpage

\section{Introduction}

Physical models and corresponding computer codes enable to evaluate nuclear reactor performances within a computation time of a few hours (see e.g. \cite{golfier09apollo,geffraye11cathare}). However, to be relevant, an optimization or a propagation of uncertainty study requires the numerical evaluations of a significant number of reactor configurations (for instance several millions in optimization \cite{hourcade11innovative}). Because of the current limitation of computing resources, these procedures can not be directly applied to computer codes. Hence, metamodels, that provide a computationally cheap approximation of the output of computer codes, are commonly used. Metamodels are constructed from a learning base of code inputs and outputs, the generation of which is called a design of experiments. In this paper, we give a detailed analysis of the metamodeling process for the Germinal V1 thermomechanical code \cite{roche00modelling}. [Note that the metamodels of the Germinal code are typically intended to be used in an optimization process applied to a sodium cooled fast reactor \cite{hourcade13core}.]

Different metamodeling methods (neural networks, Kriging and kernel methods) are analyzed and benchmarked in this paper. Furthermore, as detailed below, metamodels can not only predict computer code results but also contribute to improve the behaviour of these codes during the design of experiments.

To understand this last point it is important to highlight that it is challenging to automatically carry out several thousands of code simulations, as is typically the case in a design of experiments. Indeed current codes in nuclear engineering are complex:

\begin{itemize}
\item The code inputs and outputs are not simple scalars. For instance, it may be necessary to generate a 3D geometry and its associated meshing (see for instance \url{http://www.salome-platform.org}). 
\item It requires significant expertise to assess or anticipate the numerical validity of a calculation. Indeed, many different convergence criteria need to be taken into account. A code output file may include several indications such as ``error'' or ``warning'', whose impact is difficult to assess. 
\item There is a large number of possible calculation options that can be mixed.
\end{itemize}

\begin{figure}
\begin{center}
 \includegraphics[angle=0,width=14cm]{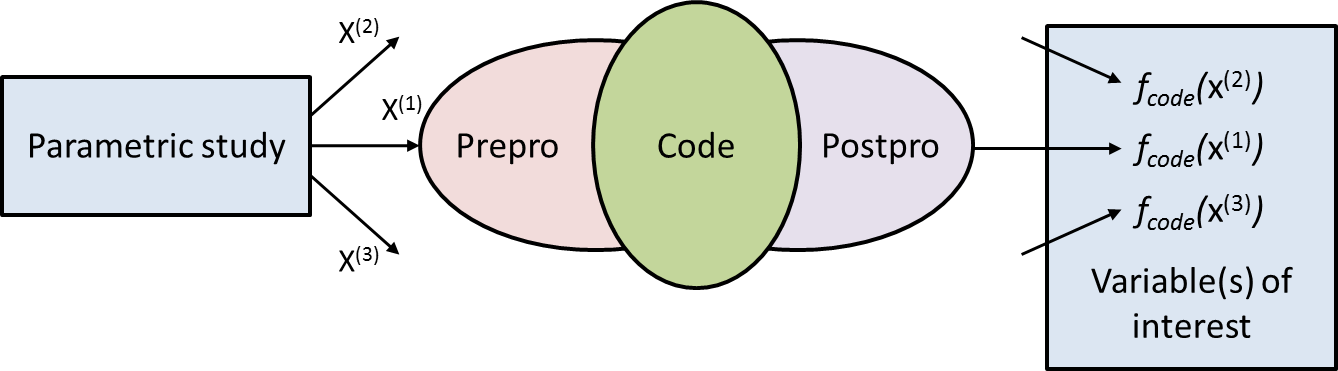}
\end{center}
\caption[]{Illustration of the code manager.
In a parametric study, all the code simulations are characterized by a finite-dimensional vector $\bx$ of scalar parameters. The code manager first consists in a preprocessor, which generates code inputs from these parameters. After the simulation, the code manager incorporates a postprocessor, which checks the occurrence of computational failure and then condense the code output in some variables of interest.}
\label{fig:code:manager}
\end{figure}

Germinal V1 is one of the multiple codes designed in the 1990s, developed to be launched manually and on a case-by-case basis. That is, for each run, an expert generates the input files and checks the consistency of the code results.
However, in a design of experiments, many code runs need to be carried out, each of which can not be managed manually.
It is hence necessary to develop a ``code manager'', as schemed in Figure \ref{fig:code:manager}. In order to explain the code manager, consider a parametric study, where each simulation is characterized by a finite number of scalar parameters (see Section \ref{section:presentation:GERMINAL} for those considered in this paper).
Then, the code manager consists first in a preprocessor which generates the code input files from the parametric variables (for example, it may automatically construct an axial mesh from a global variable like a height). After the code execution, the code manager also incorporates a postprocessor which checks the occurrence of computational failures and then condense the code output file in some variables of interest. The construction of an efficient postprocessor, able to correctly interpret all the output messages produced by the code, is a real challenge. Indeed, when many simulations are carried out for many different inputs, a very large number of failure scenarios can occur, not all of which can be anticipated.

In this paper, we consider a parametric study in which the Germinal code is used to evaluate the thermalmechanical response of a nuclear fuel pin to irradiation. Each simulation is characterized by $11$ parameters (given in Section \ref{section:presentation:GERMINAL}), and we consider a single scalar variable of interest for the simulation results. Hence, the code manager is represented by a function $f_{code}$ from $\RR^{11}$ to $\RR$ that the metamodels under consideration aim at approximating. In the sequel, we refer to this function as the code function.

We compare the metamodels obtained from neural networks, Kriging and kernel methods. We find that neural networks require the shortest computation time for metamodel evaluation, while Kriging provides the most accurate predictions. Kernel methods, and most of all Kriging, provide valuable accuracy indicators for their predictions.

We also analyze several issues related to the construction of the code manager. We show that the postprocessor can fail to detect computation failures, and that the preprocessor can generate code input files, as a function of the simulation parameters, in an inconsistent way. This preprocessor issue yields code instabilities, that is groups of computations with similar input conditions but overly different output values.

The issue with the postprocessor is, as we show, well solved by the three metamodels. Indeed, their prediction errors for the simulations in the learning base can be investigated, and a few outlier computations can be flagged. It is then possible to manually check these computations and to confirm their numerical failures.

On the other hand, code instabilities are significantly more difficult to handle and we find that Kriging provides the best tools to tackle them. Indeed, the estimated nugget effect in the Kriging metamodel (to be defined in Section \ref{subsection:kriging}) is a direct quantifier of small scale variations of the Germinal code function. This nugget effect turns out to be large in the first learning base we have considered. As a consequence, we have investigated the preprocessor behaviour, and we have found and solved an important input file generation issue. This improvement of the preprocessor results in an updated version of the code manager, from which we have generated an updated Germinal simulation base. We find that the three metamodels are more accurate for predicting the updated Germinal computations. Furthermore, the estimated nugget effect for Kriging decreases between the original and updated computations, confirming the improvement of the code manager. In light of this discussion, we believe that the estimated nugget effect of Kriging is a reliable quantifier of the global order of magnitude of the code instabilities.

The rest of the paper is organized as follows. 
In Section \ref{section:presentation:GERMINAL}, we present in details the parametric study for the Germinal code. In Section \ref{section:metamodels}, we introduce the Kriging, kernel and neural network metamodels. In Section \ref{section:original}, we discuss the prediction results of the metamodels for the original Germinal computations. In Section \ref{section:outlier:detections}, we show how the metamodels, and in particular Kriging, help in detecting outlier computations and code instabilities.
In Section \ref{section:updated}, we present the resulting prediction improvement of the metamodels for the updated computations.

Finally, it should be noted that, in this paper, we do not discuss the important problem of code validation. That is, we aim at predicting the output of the Germinal code, without assessing if the picture displayed by this output is representative of the underlying physical reality. We refer to \cite{cacuci03sensitivity,cacuci10best} for references on code validation. Remark nevertheless that constructing an accurate metamodel of a code is also useful for its validation (see e.g. \cite{higdon04combining} for the Kriging metamodel).

\section{Presentation of the parametric study for the Germinal code} \label{section:presentation:GERMINAL}

\subsection{Fuel pin thermomechanical simulation with the Germinal code}

It is well known that material properties evolve when they are submitted to high neutron flux. In particular, in fast breading reactors, irradiation can have a strong impact on fuel pin thermomechanical properties. The Germinal code V1 \cite{roche00modelling} can be used to simulate the temporal evolution of these thermo-mechanical properties, resulting from irradiation.
This code implements a simplified fuel description model based on mono-group neutron flux, power and irradiation damage distribution as well as sodium inlet temperature and mass flow per pin. In this paper, the variable of interest we focus on is the fusion margin, which is the difference between the fuel melting temperature (around $2700^ \circ$) and the maximal fuel temperature obtained throughout the Germinal simulation of the fuel life. 

We consider this variable of interest since its prediction by metamodels is particularly challenging. Indeed, the computation of the fuel temperature depends on the fuel conductivity and on the heat transfer coefficients between the fuel pellets and the cladding. These coefficients depend on the irradiation in a strong non-linear manner. Hence, in the parametric study described below, the fusion margin is definitely a non-linear function of the simulation parameters. Note also that, in a context of multi-physical optimization (neutron physics, thermal-hydraulics and thermomechanics) for fast reactor core (using the TRIAD platform \cite{hourcade13core}), we have built neural network metamodels for a large number of variables of interests of the Germinal code. We have found that the fusion margin variable was the most difficult to predict by the neural networks. 

Finally, the fusion margin is interesting in that it characterizes two very different physical regimes. When it is large and positive, the fuel pin mechanical properties do not change throughout the simulation. When, it is small, or negative, they do, which results in much more involved physical phenomena, that are challenging to model numerically. As a result, the fusion margin is generally more difficult to predict by metamodels when it is small or negative.

\subsection{The parametric study}

We are interested in a parametric study where a Germinal simulation is characterized by $11$ scalar parameters $x_1,...,x_{11}$ defined as follows. [These parameters are used by the preprocessor to generate input files for the Germinal code, see Figure \ref{fig:code:manager}.] 

\begin{figure}
\begin{center}
 \includegraphics[angle=0,width=14cm]{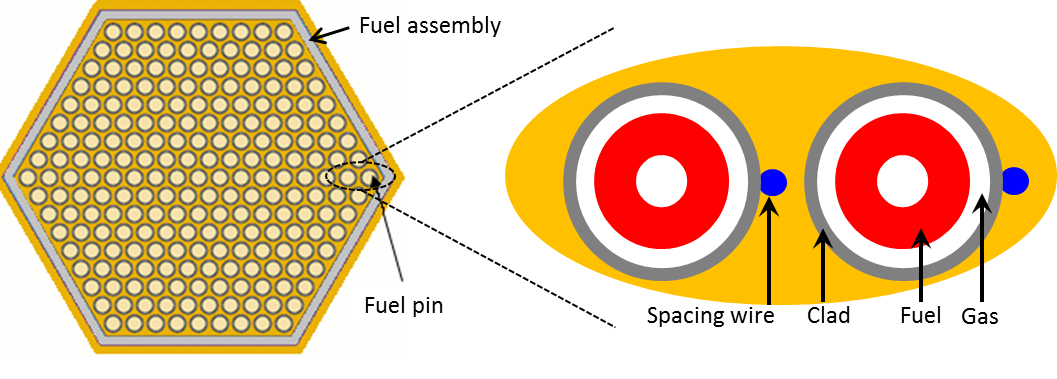}
\end{center}
\caption[]{A schematic representation of a fuel pin and a fuel assembly in nuclear fast-neutron reactors.}
\label{fig:fuel:pin}
\end{figure}

\begin{itemize}
\item The parameter $x_1$ is the cycle length in the fuel pin simulation.
\item The parameters $x_2,...,x_{7}$ characterize the nature of the fuel pin. The parameter $x_2$ is the plutonium concentration, $x_3$ is the diameter of the fuel hole, $x_4$ is the external diameter of the clad, $x_5$ is the thickness of the gap between the fuel and the clad, $x_6$ is the thickness of the clad and $x_7$ is the height of the fuel pin. Figure \ref{fig:fuel:pin} provides a visualization of $x_3$ to $x_6$.
\item The parameters $x_8,x_9,x_{10}$ characterize the power map in the fuel pin, with $x_8$ the average power, $x_9$ the axial form factor and $x_{10}$ the power shift due to the fuel depletion. [Note that, in the case of multi-physics coupling \cite{hourcade13core,lepallec11hemera}, $x_9$, $x_{10}$ and $x_{11}$ would be obtained from a neutron-physic simulation.]
\item The parameter $x_{11}$ is the volume of expansion for fission gas.
\end{itemize}

When building the learning base for the metamodels, we consider an hypercubic domain for the $11$ input variables, characterized by $11$ minimal and maximal values, summarized in Table \ref{table:input:variables}. The learning base is then obtained by first generating a LHS-Maximin \cite{santner03design} set of input parameters $\bx = (x_1,...,x_{11})$ on this hypercubic domain, and then removing some of them that can be shown to result in infeasible fuel pins prior to carrying out the corresponding Germinal simulation. The resulting learning base under consideration in this paper includes $3807$ input vectors $\bx$. We write $y_1 = f_{code}(\bx^{(1)}),...,y_n = f_{code}(\bx^{(n)})$ for the $n=3807$  computation inputs and outputs.

\begin{table}
\begin{center} 
\begin{tabular}{| c | c |  c |  c | c |  c |  c |}
\hline
parameter & $x_1$: cycle & $x_2$: plutonium & $x_3$: hole & $x_4$: external & $x_5$: fuel  & $x_6$: clad  \\
 & length & content & diameter & clad diameter & gap &  thickness \\
 & (EFPD) & ($\%$ atomic) & ($mm$) & ($mm$) & ($mm$) & ($mm$) \\
\hline
min &  360 &	10 &	0.125 &	6.2 &	0.1 &	0.5 \\
\hline
max &  440 &	30 &	3	& 12.8 &	0.2 &	0.6\\
\hline
\end{tabular}

\vspace{1cm}

\begin{tabular}{ | c |  c |  c | c |  c |  c |}
\hline
parameter & $x_7$: pin & $x_8$: average & $x_9$: axial & $x_{10}$: power & $x_{11}$: volume  \\
 & height & pin power & form factor & shift & of expansion  \\
 & ($mm$) & ($W/cm$) &  &  & ($cm^3$) \\
\hline
min & 60 &	150 &	1 &	0.8 &	32 \\
\hline
max &  160 &	440 &	1.6 &	1.2 &	94\\
\hline
\end{tabular}
\end{center} 
\caption{Minima and maxima of the intervals of variations for the $11$ simulation parameters in the parametric study. [EFPD stands for Equivalent Full Power Day.]}
\label{table:input:variables} 
\end{table}

\section{Presentation of the metamodels} \label{section:metamodels}

\subsection{Kriging} \label{subsection:kriging}

Kriging is widely used in nuclear engineering, for instance for metamodeling of computer codes \cite{lockwood12gradient} or for improving their predictions using experimental results \cite{bachoc14calibration}. In this paper, we use a standard implementation of Kriging \cite{santner03design,stein99interpolation,rasmussen06gaussian}, at the exception of the numerical optimization of the likelihood, see below.

\paragraph{Gaussian process model.}

The Kriging metamodel is based on modeling a deterministic computer code function $f_{code}: \mathcal{D} \subset \RR^d \to \RR$ (where in Section \ref{section:presentation:GERMINAL}, $d=11$) as a realization of a Gaussian process $Y$ on $\mathcal{D}$. That is, we assume, for $\bx \in \mathcal{D}$,
\begin{equation*} 
f_{code}(\bx) = Y(\omega,\bx),
\end{equation*}
where $\omega$ is a fixed element in a probability space $\Omega$. Hence, the paradigm is that, although the computer code function is fixed, the value $f_{code}(\bx)$ remains unknown to the user as long as a computation is not carried out for the conditions $\bx$. The Kriging metamodel thus follows a Bayesian approach, by considering the unknown $f_{code}(\bx)$ as the realization of a Gaussian variable $Y(\bx)$ (see also the presentation in \cite{bachoc14calibration}). 

In this paper, we assume that the Gaussian process $Y$ has a zero mean function and is hence characterized by its covariance function $C: \mathcal{D} \times \mathcal{D} \to \RR$.
In general, this covariance function is assumed to be continuous, so that the Gaussian process $Y$ yields continuous realizations, which is consistent with the fact that the code function $f_{code}$ is continuous, or at least that a small variation in the condition $\bx$ causes only a small change in the computed value $f_{code}(\bx)$. Nevertheless, the Germinal code function studied in this paper is subject to small scale variations, meaning that a small variation in $\bx$ can cause a significant change in $f_{code}(\bx)$, because of the code instabilities. Hence, we assume that
\begin{equation} \label{eq:parameterization:delta}
C(\bx^{(a)},\bx^{(b)}) = \sigma_0^2 \bar{C}(\bx^{(a)},\bx^{(b)}) + \delta_0^2 \mathbf{1} \{\bx^{(a)} = \bx^{(b)} \},
\end{equation}
with $\mathbf{1} \{.\}$ the indicator function, where $\bar{C}$ is a continuous correlation function and with $\sigma_0^2 > 0$ and $\delta_0^2 \geq 0$. Thus, $Y$ can be written
\begin{equation} \label{eq:continuous:discontinuous}
Y(\bx) = Y_{c}(\bx)  + Y_d(\bx),
\end{equation}
where $Y_c(\bx)$ is a continuous Gaussian process with covariance function $\sigma_0^2 \bar{C}(\bx^{(a)},\bx^{(b)})$
and $Y_d(\bx)$ is a discontinuous Gaussian process with covariance function $\delta_0^2 \mathbf{1} \{\bx^{(a)} = \bx^{(b)} \}$. The term $\delta_0^2 \mathbf{1} \{\bx^{(a)} = \bx^{(b)} \}$ is referred to as a nugget effect \cite{andrianakis12effect}.

\paragraph{Covariance parameter estimation.}

Most classically in Kriging, the covariance function $C$ is estimated, from the computations $y_1 = f_{code}(\bx^{(1)}),...,y_n = f_{code}(\bx^{(n)})$ in the learning base. In this paper, we estimate the covariance function within the parametric set
\begin{equation} \label{eq:parameterization:alpha}
\mathcal{C} = \left\{ \sigma^2 \left[ \bar{C}_{\bell}( \bx^{(a)} - \bx^{(b)} ) + \alpha \mathbf{1} \{\bx^{(a)} = \bx^{(b)} \} \right] , \sigma >0 , \bell \in (0,+\infty)^d , \alpha \geq 0 \right\},   
\end{equation} 
where $\bar{C}_{\bell}( \bh )$ is the Mat\'ern $3/2$ correlation function,
\[
\bar{C}_{\bell}( \bh) = ( 1 + \sqrt{6} |\bh|_{\bell} )  \exp{ ( - \sqrt{6} |\bh|_{\bell} ) }, 
\]
with $|\bh|_{\bell} = \sqrt{ \sum_{i=1}^d \frac{h_i^2}{\ell_i^2} }$. 

The Mat\'ern $3/2$ correlation function $ \bar{C}_{\bell}( \bh )$ is one of the most commonly used covariance functions. It is stationary, that is $\bar{C}_{\bell}( \bx^{(a)} - \bx^{(b)} )$ depends on $\bx^{(a)}$ and $\bx^{(b)}$ only through their difference. Furthermore, for every $\bell$, this correlation function yields Gaussian process realizations that are exactly one time continuously differentiable (see for instance \cite{stein99interpolation}).
The component $\ell_i$ can be seen as a correlation length in the $i$-th dimension. When $\ell_i$ is small, the condition $x_i$ is particularly important for the Gaussian process $Y(\bx)$. Conversely, if $\ell_i$ is very large, then the realizations of $Y(\bx)$ are almost independent of $x_i$.

The covariance parameters $\sigma^2, \bell,\alpha$ are estimated from the learning base. In this paper, we address Maximum Likelihood estimation (ML) which is the most standard method. [We note that other methods can also be employed, like Cross Validation \cite{rasmussen06gaussian,bachoc13cross,bachoc14asymptotic}.] Let, $\R_{\bell,\alpha}$ be the $n \times n$  matrix defined by $(R_{\bell,\alpha})_{i,j} =  \bar{C}_{\bell}( \bx^{(i)} - \bx^{(j)} ) + \alpha \mathbf{1} \{\bx^{(i)} = \bx^{(j)} \}$. Let $\by$ be the $n \times 1$ vector $(y_1,...,y_n)^t$. Then, the Maximum Likelihood estimator of $\sigma^2,\bell,\alpha$ is defined by

\begin{equation} \label{eq:ML:un}
(\hat{\bell} , \hat{\alpha}) \in \argmin_{(\bell,\alpha)}  \log \left( \frac{1}{n} \by^t \R_{\bell,\alpha}^{-1} \by  \right) + \frac{1}{n} \log( |\R_{\bell,\alpha}| ),
\end{equation}
where $|.|$ is the determinant, and by 
\begin{equation} \label{eq:ML:deux}
\hat{\sigma}^2 = \frac{1}{n} \by^t \R_{\hat{\bell},\hat{\alpha}}^{-1} \by.
\end{equation}

The nugget variance $\delta_0^2$ in \eqref{eq:parameterization:delta} is estimated by $\hat{\delta}^2 = \hat{\sigma}^2 \hat{\alpha}$. Note that the advantage of the parameterization with $\sigma^2,\bell,\alpha$ in \eqref{eq:parameterization:alpha}, compared to a parameterization with $\sigma^2,\bell,\delta^2$ as in \eqref{eq:parameterization:delta}, is that it provides an explicit expression for $\hat{\sigma}^2$. 

\paragraph{Numerical optimization of the likelihood.}

The optimization problem \eqref{eq:ML:un} is relatively challenging in our case, since the optimization space has dimension $d+1 = 12$ and since $n$ is around $3800$ which makes it computationally costly to evaluate the determinant and to solve the linear system in \eqref{eq:ML:un}.

Hence, we evaluate $\hat{\bell}$ and  $\hat{\alpha}$ in two steps. First, we select a random subsample of the
learning base, of size $1000$ and minimize the equivalent of the function \eqref{eq:ML:un}, when the learning base is equal to this random subsample. We let $\hat{\bell}$ and $\tilde{\alpha}$ be the outcome of this first step.

For the second, step, recall that $\R_{\hat{\bell},\tilde{\alpha}}$ is the $n \times n$  matrix defined by $(R_{\hat{\bell},\tilde{\alpha}})_{i,j} =  \bar{C}_{\hat{\bell}}( \bx^{(i)} - \bx^{(j)} ) + \tilde{\alpha} \mathbf{1} \{\bx^{(i)} = \bx^{(j)} \}$. Consider a SVD decomposition of this matrix, $\R_{\hat{\bell},\tilde{\alpha}} = \U \S \U^t$, with $\U$ of size $n \times n$ so that $\U \U^t = \I_n$ and $\S$ a diagonal matrix with diagonal elements $ s_1 \geq... \geq s_n >0$. Then, let $L_{\alpha}$ be the function in \eqref{eq:ML:un} evaluated at $\hat{\bell},\alpha$. This function can be written, with $v_i = ( \U^t \by )_i$,
\begin{equation} \label{trick:svd:lambda}
L_{\alpha} = \log \left( \frac{1}{n} \sum_{i=1}^n \frac{v_i^2}{ s_i +   \alpha -\tilde{\alpha}  } \right) + \frac{1}{n} \sum_{i=1}^n \log \left( s_i  + \alpha  - \tilde{\alpha}  \right),
\end{equation}
which is computed with negligible computational cost. Hence, we can plot the graph of $L_{\alpha}$ and compute $\hat{\alpha}$ as its minimizer. Finally, $\hat{\sigma}^2$ is computed by \eqref{eq:ML:deux}.

Hence, in this two-step optimization procedure, only the first step entails an important computational cost. The second step is carried out in negligible time and provides an estimation of the nugget component $\alpha$ that is more accurate since all the elements of the learning base are used.

\paragraph{Prediction.}

Once the estimators $\hat{\sigma}^2, \hat{\bell},\hat{\alpha}$ are computed, the standard ``plug-in'' approach \cite{stein99interpolation} is to assume that the covariance function is known and equal to that obtained from the estimators, that is 
\[
C(\bx^{(a)},\bx^{(b)}) = 
\hat{\sigma}^2 \left[ \bar{C}_{\hat{\bell}}( \bx^{(a)} - \bx^{(b)} ) + \hat{\alpha} \mathbf{1} \{\bx^{(a)} = \bx^{(b)} \} \right]. 
\]
We make this assumption, which enables to construct the Kriging metamodel of $f_{code}$ as follows. Let $\R$ be a shorthand for $\R_{\hat{\bell},\hat{\alpha}}$. Let, for $\bx \in \mathcal{D}$, $\br(\bx)$ be the $n \times 1$ vector defined by $(r(\bx))_i = \bar{C}_{\hat{\bell}}( \bx - \bx^{(i)} ) + \hat{\alpha} \mathbf{1} \{\bx = \bx^{(i)} \}$. Then, conditionally to $\by$, $Y(\bx)$ follows a Gaussian distribution with mean
\begin{equation} \label{eq:hatf:kriging}
\hat{f}_{code}(\bx) = \br(\bx)^t \R^{-1} \by,
\end{equation}
and variance
\begin{equation} \label{eq:predictive:variance:krig}
\hat{\sigma}^2_{code}(\bx) = \hat{\sigma}^2 \left( 1 + \hat{\alpha} - \br(\bx)^t \R^{-1} \br(\bx) \right).
\end{equation}

In the above display, $\hat{f}_{code}(\bx)$ is the metamodel function of $f_{code}$, that can be compared with those obtained from the artificial neural network and kernel regression methods. The quantity $\hat{\sigma}^2_{code}(\bx)$, that we call the predictive variance, is however specific to Kriging. It is one of the benefits of considering a Gaussian process model for $f_{code}$. The predictive variance can be used, for instance, to construct the confidence interval $[ \hat{f}_{code}(\bx) - 1.65 \hat{\sigma}_{code}(\bx) , \hat{f}_{code}(\bx) + 1.65 \hat{\sigma}_{code}(\bx) ]$ that contains $Y(\bx)$ with probability $0.9$. 

Note that, for any $\bx$ which does not belong to $\{ \bx^{(1)},...,\bx^{(n)} \}$, we have with the notation of \eqref{eq:continuous:discontinuous}, and where $\EE$ denotes the expected value,
\begin{equation} \label{eq:interpretation:nugget:prediction}
\hat{\sigma}^2_{code}(\bx) = \EE \left(  (\hat{f}_{code}(\bx) - Y_c(\bx) )^2  \right) + \EE( Y_d(\bx)^2 )  = \EE \left(  (\hat{f}_{code}(\bx) - Y_c(\bx) )^2 \right) + \hat{\delta}^2.
\end{equation}

This is interpreted as follows: The value of the discontinuous Gaussian process $Y_d(\bx)$ in \eqref{eq:continuous:discontinuous} can not be inferred from the values of $Y_d(\bx^{(1)}),...,Y_d(\bx^{(n)})$ (predicting $Y_d(\bx)$ by $0$ is in fact the best possibility).
Hence, the prediction mean square error $\hat{\sigma}^2_{code}(\bx)$ for $Y(\bx)$ is larger than $\hat{\delta}^2$, and the difference between $\hat{\sigma}^2_{code}(\bx)$ and $\hat{\delta}^2$ corresponds to the prediction error for $Y_c(\bx)$, which is the continuous component of $Y(\bx)$. Thus, in practice, the square prediction error for the code function $f_{code}(\bx)$ should be on average larger than $\hat{\delta}^2$.

We conclude this presentation of Kriging with the virtual cross validation formulas \cite{bachoc13cross,dubrule83cross}. Consider $\hat{\sigma}^2,\hat{\bell},\hat{\alpha}$ to be estimated from the learning base $y_1 = f_{code}(\bx^{(1)}),...,y_n = f_{code}(\bx^{(n)})$ and fixed. Then, let $\hat{f}_{code,LOO}(\bx^{(i)})$ and $\hat{\sigma}^2_{code,LOO}(\bx^{(i)})$
be the Leave-One-Out (LOO) prediction and predictive variance for $f_{code}(\bx^{(i)})$, that would be obtained from \eqref{eq:hatf:kriging} and \eqref{eq:predictive:variance:krig} if $\bx^{(i)}$ and $f_{code}(\bx^{(i)})$ were removed from the learning base. Then we have, for $1 \leq i \leq n$,
\begin{equation} \label{eq:Kriging:LOO:prediction}
f_{code}(\bx^{(i)}) - \hat{f}_{code,LOO}(\bx^{(i)}) = \frac{1}{ ( \R^{-1} )_{i,i} } ( \R^{-1} \by )_i
\end{equation}
and
\begin{equation} \label{eq:Kriging:LOO:predictive:variance}
\hat{\sigma}^2_{code,LOO}(\bx^{(i)}) = \frac{1}{ ( \R^{-1} )_{i,i} }.
\end{equation}
Hence, the $n$ LOO errors and predictive variances can be computed by means of a single $n \times n$ matrix inversion, while a naive approach, consisting in evaluating $n$ different versions of \eqref{eq:hatf:kriging} and \eqref{eq:predictive:variance:krig}, would necessitate to solve $n$ linear systems of size $(n-1) \times (n-1)$.

\subsection{Kernel methods}

Kernel methods \cite{wahba90Spline,scholkopf02learning} are frequently used for statistical learning
and metamodeling. The kernel metamodel eventually yields prediction formula similar to Kriging (compare \eqref{eq:hatf:kriging} and \eqref{eq:hatf:kernel}), although the philosophy is different.

Kernel methods, for inputs in a domain $\mathcal{D} \subset \mathbb{R}^d$, are based on a symmetric nonnegative definite kernel function $k : (\bx,\by) \in \mathcal{D}^2 \rightarrow \R$, see \cite{scholkopf02learning}. This kernel function defines a Hilbert space $\mathcal{H}_k$ of functions from $\mathcal{D}$ to $\mathbb{R}$, that is called the Reproducing Kernel Hilbert Space (RKHS) corresponding to the kernel function (see \cite{scholkopf02learning} for details). 

Consider now the learning base $(\bx^{(1)},y_1=f_{code}(\bx^{(1)})),...,(\bx^{(n)},y_n=f_{code}(\bx^{(n)}))$.
Then, for each $\lambda \geq 0$, that we call the regularity parameter, we can consider the function $\hat{f}_{\lambda} \in \mathcal{H}_k$  which solves
\begin{eqnarray} \label{eq:abstract:optim:problem}
\hat{f}_{\lambda} = \argmin_{f \in \mathcal{H}_k} \frac{1}{n} \sum_{i=1}^n (y_i - f(\bx_i))^2 + \lambda ||f||_{\mathcal{H}_k}^2,
\end{eqnarray}
where $||f||_{\mathcal{H}_k}$ is a complexity measure for the function $f$, see \cite{scholkopf02learning}. Thus, the aim is that the function $\hat{f}_{\lambda}$ both reproduce well the observations $y_i$ and be of small complexity, in order to prevent overfitting. Increasing the value of $\lambda$ prevents overfitting all the more.

It turns out that the abstract optimization problem \eqref{eq:abstract:optim:problem} has an explicit solution that is computable in practice. Let $\R_{\lambda}$ be the $n \times n$ matrix defined by $(\R_{\lambda})_{i,j} = k(\bx^{(i)},\bx^{(j)}) + n \lambda \mathbf{1} \{i=j\}$, let $\br(\bx)$ be the $n \times 1$ vector defined by $r(\bx)_i = k(\bx^{(i)},\bx)$ and let $\by = (y_1,...,y_n)^t$. Then, we have
\begin{eqnarray} \label{eq:hatf:kernel}
\hat{f}_{\lambda}(\bx) = \boldsymbol{r}(\bx)^t  \R_{\lambda}^{-1} \by.
\end{eqnarray}

Note that, when $\lambda=0$ and $\R_0$ is of full rank, we obtain an exact interpolation: $\hat{f}_{0}(\bx^{(i)}) = y_i$. Nevertheless, using a non-zero $\lambda$
enables us to deal with the small scale variations of the Germinal code (similarly to the nugget effect of the Kriging metamodel). Note also that calculating  $\R_{\lambda}^{-1} \by $ is more convenient numerically when $\lambda$ is large. 

We select the value of the regularity parameter $\lambda$ by
Generalized Cross Validation (GCV) \cite{golub79generalized}. 
The selected $\lambda$ is given by
\begin{eqnarray} \label{eq:lambda:GCV}
\lambda_{GCV} = \arg \min_{\lambda}  \frac{|| \R_{\lambda}^{-1} \by ||}{\texttt{Trace}( \R_{\lambda}^{-1})},
\end{eqnarray}
where $||.||$ is the Euclidean norm. Hence, the final kernel metamodel function is
$\hat{f}_{code} = \hat{f}_{\lambda_{GCV}}$. Note that the minimization problem \eqref{eq:lambda:GCV} entails a negligible computation cost, since a SVD decomposition can be used, similarly to \eqref{trick:svd:lambda} for Kriging. 

Because the prediction formula \eqref{eq:hatf:kriging} and \eqref{eq:hatf:kernel} are identical, there exist virtual LOO formulas for kernel methods, that are similar to those of Kriging in \eqref{eq:Kriging:LOO:prediction}. By letting $\R$ be $\R_{\lambda_{GCV}}$, we have
\begin{eqnarray}
f_{code}(\bx^{(i)}) - \hat{f}_{code,LOO}(\bx^{(i)}) = \frac{1}{ ( \R^{-1} )_{i,i} } ( \R^{-1} \by )_i,
\end{eqnarray}
where $\hat{f}_{code,LOO}(\bx^{(i)})$ is defined as in \eqref{eq:Kriging:LOO:prediction} but for the kernel metamodel. 

Note that, since kernel methods are not based on a probabilistic model, there are no error indicators similar to $\hat{\sigma}_{code}^2 (\bx)$ in \eqref{eq:predictive:variance:krig} for Kriging.

In this paper, the kernel function $k$ we consider is defined by $k(\bx,\by)=\prod_{i=1}^d \bar{k}(x_i,y_i)$ with
\begin{eqnarray}
\bar{k}(x,y)	= \sum_{l=0}^m \frac{1}{(l!)^2} B_l(x)B_l(y) + \frac{(-1)^{m+1}}{(2m)!} B_{2m}(|x-y|),
\end{eqnarray}
where $B_l$ is the $l$-th Bernoulli polynomial. The benefit of this kernel function is that the corresponding RKHS $\mathcal{H}_k$ consists in the Sobolev space of functions that are $m$ times differentiable \cite{wahba90Spline,yue99Doe}. Hence $m$ can be chosen according to the smoothness we require from the metamodel function.
We choose $m=2$ in this paper, as it provides the minimal value for \eqref{eq:lambda:GCV}.

\subsection{Artificial neural networks} \label{subsection:rn}

Artificial Neural Networks (ANNs) are known as efficient modelling tools to approximate nonlinear functions with the fundamental property of parsimonious approximation \cite{dreyfus05neural}. We carried out all computations for the neural networks with the uncertainty quantification platform URANIE \cite{gaudier10uranie}. We consider the Multi Layer Perceptron (MLP) \cite{bishop95neuron} with one hidden layer and one output. The MLP consists of simple connections between neurons and is characterized by the number of hidden neurons and the weights of their corresponding connections.

For a given number of hidden neurons, the weights are fitted by using the standard back-propagation procedure \cite{rumelhart86learning}. This procedure is repeated with different weight initializations, and the eventual values of the weights are selected by cross validation. Finally, the number of hidden neurons is selected by a minimization of the RMSE (Root Mean Square Error) on the full learning data set.

\subsection{Computation times}

The Kriging and neural network metamodels are used in two steps. First, in what we call the construction phase, the neural network structure and the covariance parameters for Kriging are optimized (Sections \ref{subsection:kriging} and \ref{subsection:rn}). This first step yields the metamodel function $\hat{f}_{code}$. Note that, beneficially, the construction phase is not needed for kernel regression.
Second, in what we call the evaluation phase, for many inputs $\bx$, the metamodel predictions $\hat{f}_{code}(\bx)$ are calculated. 

With the implementation we used, for the learning base under consideration, and on a personal computer, the computation time for the construction phase is around five to ten hours for both Kriging and neural networks. Since this typically takes place only once, this time is not critical. The evaluation time is, on average, $0.00015$ seconds per input $\bx$ for the neural networks and $0.004$ seconds for Kriging and kernel methods. Hence, neural network evaluation is faster, which is explained because the evaluation cost of the neural network metamodel function is proportional to the number of hidden neurons, while those of the Kriging or kernel regression metamodel functions are proportional to $n$. In our case, $n$ is much larger than the number of hidden neurons. For the three metamodels, the evaluation times are not prohibitive for using the metamodel functions in an optimization framework, like in \cite{hourcade13core}, where a few millions of metamodel evaluations would be required.

\section{Prediction and classification results for the original Germinal computations} \label{section:original}

\subsection{Prediction results} \label{subsection:prediction:results}

\paragraph{Estimated covariance parameters for Kriging.}
The estimated covariance parameters for the Kriging metamodel are presented in Table \ref{table:Germinal:estimation:MF}. Note that we have applied an affine standardization of $\{\bx^{(1)},...,\bx^{(n)}\}$ in $[0,1]^{11}$, to obtain, for $i=1,...,11$, $\min_j \bx^{(j)}_i = 0$ and $\max_j \bx^{(j)}_i = 1$. Hence, for the correlation length vector $\bell$ we present, all the components are at the same scale and should be compared to inputs in $[0,1]^{11}$.

\begin{table}
\begin{center} 
\begin{tabular}{| c | c |  c |  c | c |  c |  c | c |  c |  c | c |  c |  c | c |}
\hline
$\hat{\sigma} (^{\circ})$ &  $\hat{\ell}_1$ & $\hat{\ell}_2$ &$\hat{\ell}_3$ &$\hat{\ell}_4$ &$\hat{\ell}_5$ &$\hat{\ell}_6$ &$\hat{\ell}_7$ &$\hat{\ell}_8$ &$\hat{\ell}_9$ &$\hat{\ell}_{10}$ &$\hat{\ell}_{11}$ & $\hat{\delta} (^{\circ})$ \\ 
\hline
 $1264$  &  $21$   & $50$  &  $12$ &  $4.5$  &  $12$   & $64$  &  $100$  &  $2.2$  &  $6.6$  &  $5.9$  &  $100$ &  $28.5$   \\
\hline
\end{tabular}
\end{center} 
\caption[Estimated covariance parameters for Kriging.]{Estimated covariance parameters $(\hat{\sigma},\hat{\bell},\hat{\delta})$ for the Kriging metamodel of the fusion margin output of the Germinal code.}
\label{table:Germinal:estimation:MF} 
\end{table}

The input variables $x_i$ with smallest estimated correlation lengths $\hat{\ell}_i$ are considered the most influential for the code function in the Kriging model. In Table \ref{table:Germinal:estimation:MF}, the smallest estimated correlation length is $\hat{\ell}_8$, corresponding to the average pin power input.
This is natural, since the average pin power has a strong direct influence on the power map in the fuel pin, which is intrinsically related to the temperature in the fuel pin and thus to the fusion margin.
Similarly, the inputs $x_{9}$ (axial form factor) and $x_{10}$ (power shift) impact the power map and the inputs $x_3$ (hole diameter) and $x_4$ (external clad diameter) characterize the geometry of the fuel pin. These four inputs thus have a strong impact on the fusion margin, so that their corresponding estimated correlation lengths are also relatively small.

On the contrary, in the Kriging model, the code output is considered unaffected by the values of the input variables $x_i$ with very large correlation lengths. [Note that in Table \ref{table:Germinal:estimation:MF}, the maximum correlation length value is $100$, which is the upper bound in the likelihood optimization procedure, and is practically equivalent to an infinite correlation length.] The two input variables with correlation lengths $100$ are $x_{7}$ (pin height) and $x_{11}$ (volume of expansion).
Indeed, the fuel power (and thus the temperature) is not related to the pin height. Furthermore, the volume of expansion has no physical link with the temperature.
 
The estimated nugget variance is $\hat{\delta}^2 = (28.5^{\circ})^2$, which is a signal that code instabilities might be present, as confirmed in Section \ref{subsection:reduction:nugget:effect}, and which indicates that the RMSE should be at least around $30 ^{\circ} $, as is confirmed below. This interpretation of the covariance parameters of Kriging is hence beneficial and constitutes and asset, in comparison with neural networks and kernel methods.

\paragraph{Prediction criteria.}

We evaluate the accuracy of the metamodels by using a test base $(\bx_t^{(1)},f_{code}(\bx_t^{(1)})),...,(\bx_t^{(n_t)},f_{code}(\bx_t^{(n_t)}))$, that is generated independently from and in the same way as the learning base, with $n_t = 1613$.

The first criterion we consider is the Root Mean Square Error on the test base (RMSE), with
$\hat{f}_{code}(\bx)$ the prediction of $f_{code}(\bx)$, obtained from the artificial neural network, Kriging or kernel methods,
\begin{equation} \label{eq:RMSEt}
 RMSE^2 =  \frac{1}{n_t} \sum_{i=1}^{n_t} \left( \hat{f}_{code}(\bx_t^{(i)}) - f_{code}(\bx_t^{(i)}) \right)^2.  
\end{equation}

A second criterion is the $Q^2$ (considered for instance in \cite{marrel08efficient}) defined by
\begin{equation} \label{eq:Qdeux}
 Q^2 =  1 - \frac{RMSE^2}{sd_{code}^2},  
\end{equation}
where $sd_{code}$ is the standard deviation of the output on the test base (the standard deviation of $\{ f_{code}(\bx_t^{(1)}),...,f_{code}(\bx_t^{(n_t)}) \}$). The $Q^2$ is thus a relative efficiency criterion, whose value is always smaller than $1$ and increases with the accuracy of the predictions.

The criteria $RMSE$ and $Q^2$ are not observable in practice, but can be estimated from the learning base. In order to do so, let $\tilde{f}_{code}(\bx^{(i)})$ be the prediction $\hat{f}_{code}(\bx^{(i)})$ of $f_{code}(\bx^{(i)})$ obtained from the artificial neural network, or the LOO prediction $\hat{f}_{code,LOO}(\bx^{(i)})$ of $f_{code}(\bx^{(i)})$ with Kriging or kernel methods. Then, $RMSE$ and $Q^2$ can be estimated by $\widehat{RMSE}$ and $\widehat{Q}^2$, defined by
\begin{equation} \label{eq:RMSEl}
 \widehat{RMSE}^2 =  \frac{1}{n} \sum_{i=1}^{n} \left( \tilde{f}_{code}(\bx^{(i)}) - f_{code}(\bx^{(i)}) \right)^2  
\end{equation}
and
\begin{equation} \label{eq:Qdeuxl}
 \widehat{Q}^2 = 1 - \frac{\widehat{RMSE}^2}{\widehat{sd}_{code}^2},  
\end{equation}
where $\widehat{sd}_{code}$ is the standard deviation of the output on the learning base.

Then, for $\gamma \in (0,1)$, we define the criterion $q_{\gamma}$ as the empirical quantile $\gamma$ of the set of errors $\left| \hat{f}_{code}(\bx_t^{(i)}) - f_{code}(\bx_t^{(i)}) \right|$, for $i=1,...,n_t$.

Finally, one specificity of Kriging is that it provides the predictive variance \eqref{eq:predictive:variance:krig} which enables to build predictive confidence intervals for the code values $f_{code}(\bx)$.
To assess the accuracy of the $90\%$-confidence intervals presented after \eqref{eq:predictive:variance:krig}, we consider the following Confidence Interval Ratio (CIR), defined as
\begin{equation} \label{eq:Germinal:CIR}
 CIR =  \frac{1}{n_t} \sum_{i=1}^{n_t} \mathbf{1} \{  | \hat{f}_{code}(\bx_t^{(i)}) - f_{code}(\bx_t^{(i)}) | \leq 1.64 \hat{\sigma}_{code}(\bx_t^{(i)}) \}. 
\end{equation}
The CIR criterion is specific to Kriging and should be close to $0.9$.

\paragraph{Prediction results.}

The prediction results are given in Table \ref{table:prediction:results:MF}.
The standard deviation of the output on the test base is $sd_{code} = 342 ^{\circ}$, and the RMSE for the neural network, Kriging and kernel methods are respectively $38.5 ^{\circ}$, $36.1 ^{\circ}$ and $44.5 ^{\circ}$. The relative prediction errors are thus around $10\%$, which is a good performance considering the complexity of the fusion margin output. Similarly, the relative efficiency criteria $Q^2$ are around $99 \%$ for the three metamodels. Kriging provides slightly more accurate predictions than the neural networks, and these two metamodels perform better than kernel methods. The same hierarchy holds when we consider the quantiles $q_{0.9}$ and $q_{0.95}$ of the absolute prediction errors.

The Kriging estimate of the nugget variance is $\hat{\delta}^2 = (28.5 ^{\circ})^2$. As is seen in Section \ref{subsection:kriging}, under the Gaussian process assumption of Kriging, this value corresponds to the irreducible prediction error for $f_{code}(\bx)$, stemming from the small scale variations of $f_{code}$ which are due to code instabilities. Hence, a large part of the prediction errors of the metamodels comes from these code instabilities.

For Kriging and kernel methods, $\widehat{RMSE}$ is a very reliable estimate of RMSE, while $\widehat{RMSE}$ is moderately too optimistic for the neural networks, as it is smaller than RMSE. Indeed, the neural network functions are optimized according to their prediction errors on the learning base, so that these errors are eventually slightly smaller than the new errors on the test base. For Kriging and kernel methods, the LOO precisely avoids this phenomenon, by providing prediction errors for outputs $f_{code}(\bx^{(i)})$ that are removed from the learning base.
The $90\%$ confidence intervals provided by Kriging are also appropriate, as they contain $89.8\%$ of the output values in the test base (CIR = $89.8\%$ in \eqref{eq:Germinal:CIR}).

\begin{table}
\begin{center} 
\begin{tabular}{|c | c  |  c | c | c | c | c |}
\hline 
 & $\widehat{RMSE}$ & RMSE & $\widehat{Q}^2$ & $Q^2$ & $q_{0.9}$ & $q_{0.95}$  \\
\hline
Neural network &  $34.5 ^{\circ}$  & $38.5 ^{\circ}$ & $0.990$ & $0.987$ & $61.6 ^{\circ}$ & $76.7 ^{\circ}$ \\ 
\hline
Kriging &  $35.6^{\circ}$ &  $36.1 ^{\circ}$ &      $0.989$ & $0.989$ & $57.4 ^{\circ}$ & $72.7 ^{\circ}$ \\
\hline
Kernel methods  & $44.3 ^{\circ}$ &  $44.5 ^{\circ}$ & $0.983$ & $0.983$ &  $68.5 ^{\circ}$ & $88.8 ^{\circ}$   \\
\hline
\end{tabular}
\end{center} 
\caption[Prediction results for the fusion margin output]{Prediction results for the fusion margin output of the Germinal code (original computations). The standard deviation of the output on the test base is $342 ^{\circ}$. The quantities $RMSE$ and $Q^2$ are error and efficiency criteria for prediction on the test base. They are estimated by $\widehat{RMSE}$ and $\widehat{Q}^2$ that use the learning base, see \eqref{eq:RMSEl} and \eqref{eq:Qdeuxl}. The estimates $\widehat{RMSE}$ and $\widehat{Q}^2$ are more accurate for Kriging and kernel methods than for the neural networks, thanks to the virtual LOO formulas. 
}
\label{table:prediction:results:MF} 
\end{table}

\begin{figure}
\begin{center}
\begin{tabular}{ccc}
\hspace{-1cm} \includegraphics[angle=0,width=5cm,height=6cm]{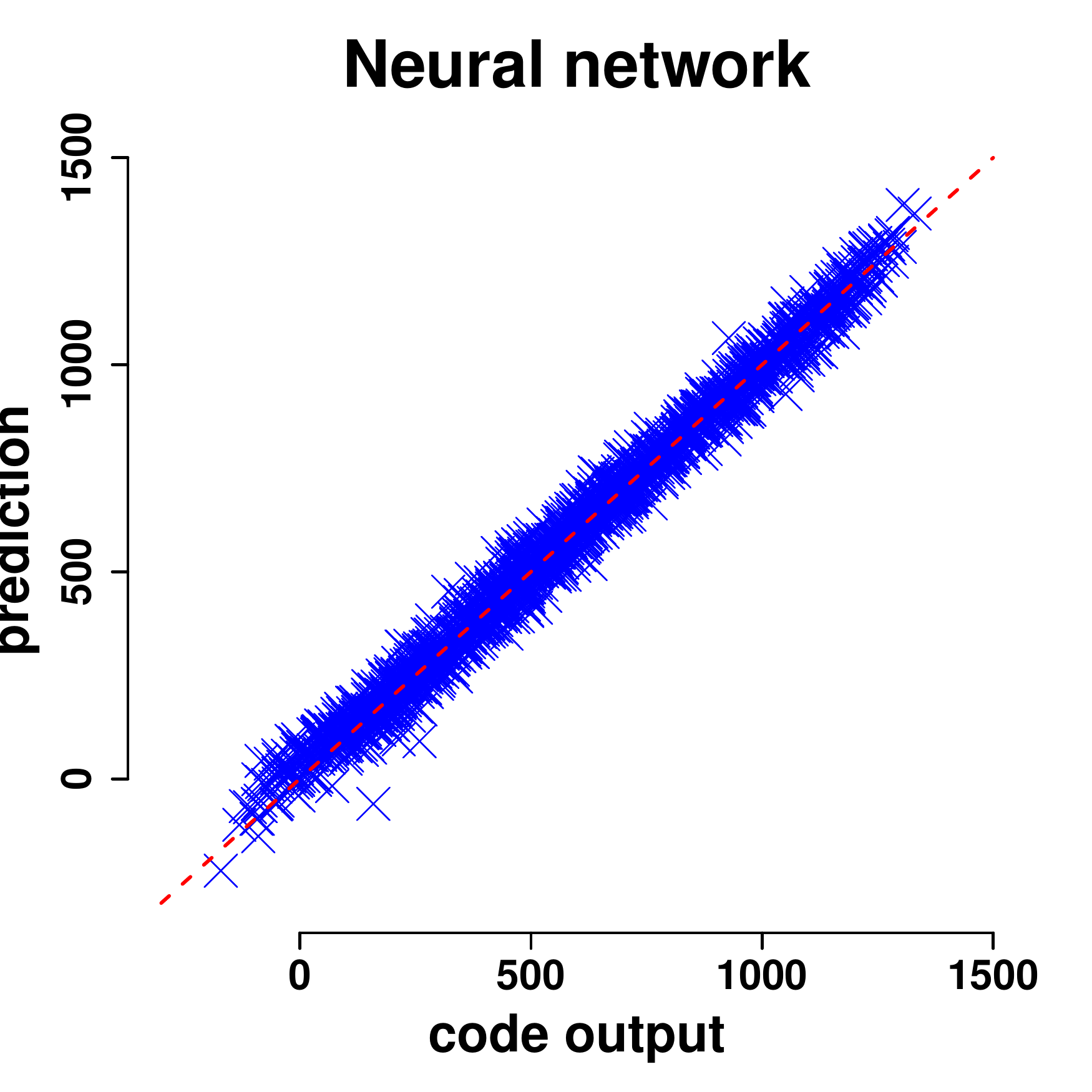}
&
\includegraphics[angle=0,width=5cm,height=6cm]{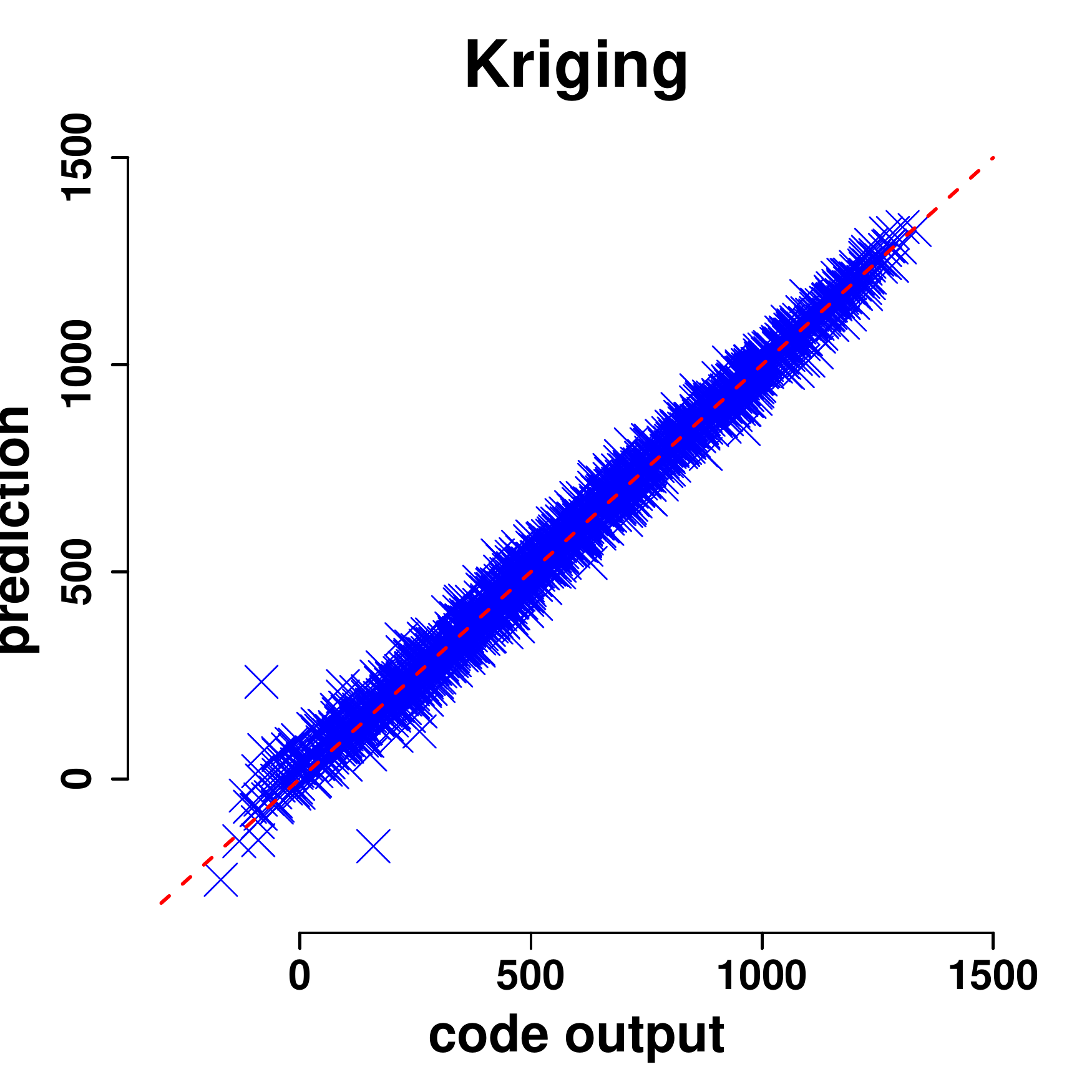}
&
\includegraphics[angle=0,width=5cm,height=6cm]{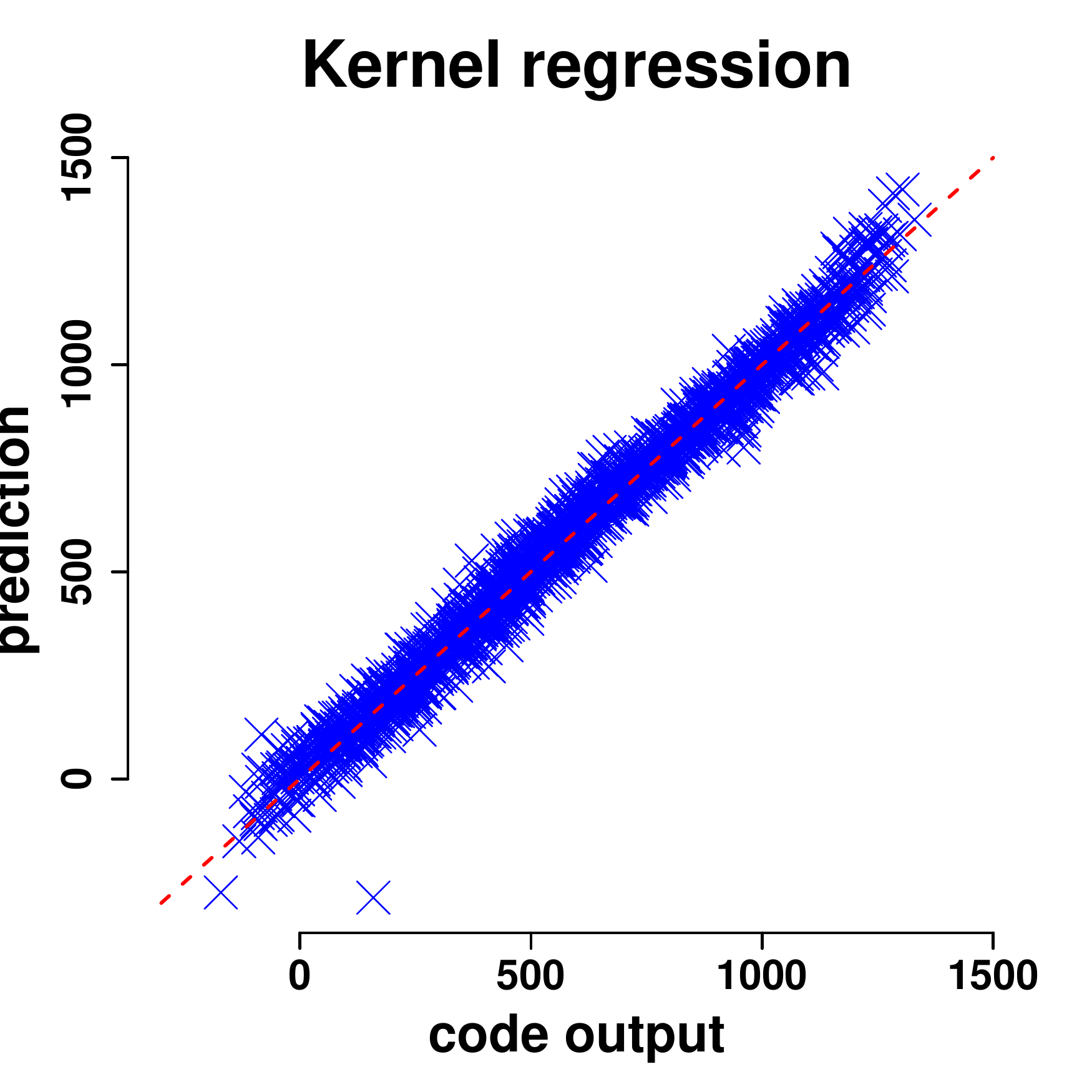}
\end{tabular}
\end{center}
\caption[Prediction errors of the fusion margin output (original computations)]{Plot of the metamodel predictions in the test base (y-axis), as a function of the Germinal output values (x-axis), for the neural networks (left), Kriging (middle) and kernel methods (right). Case of the original computations. The dashed lines is defined by $y=x$.}
\label{fig:pred:old:MF}
\end{figure}

In Figure \ref{fig:pred:old:MF}, we plot the predictions as functions of the Germinal output values. For the three metamodels, the predictions are less accurate when the fusion margin is negative or close to negative. Indeed, this corresponds to complex physical processes, that are challenging to simulate numerically, as discussed in Section \ref{section:presentation:GERMINAL}. Some of the prediction errors are particularly large and stand out in Figure \ref{fig:pred:old:MF}. We call outliers these Germinal computations that are poorly predicted, and give more comments on their detection in the learning base and their analysis in section \ref{section:outlier:detections}.

\subsection{Classification} \label{subsection:classification:old}

\paragraph{Classification goal and classifiers.}

In practice, a simulated fuel pin (characterized by $\bx$) is considered viable if the fusion margin output ($f_{code}(\bx)$) is larger than $300^ \circ$. This value of $300^ \circ$ is a security margin, accounting for the possible discrepancies between a Germinal simulation and a real utilization of a fuel pin.

Hence, besides predicting the fusion margin output $f_{code}(\bx)$, it is desirable to classify the inputs $\bx$ in two classes: Those which are viable ($f_{code}(\bx) > 300 ^ \circ$) and those which are unsafe ($f_{code}(\bx) \leq 300 ^ \circ$). Furthermore, the two possible corresponding classification errors do not have the same impact, so that it is very beneficial to have a tunable classifier, that can for example decrease the number of unsafe $\bx$ that are classified as viable, at the cost of increasing the number of viable $\bx$ that are classified as unsafe.

This tuning can be achieved naturally with metamodels. Let $\hat{f}_{code}(\bx)$ be the metamodel prediction at $\bx$ and let $\hat{\sigma}_{code}^2(\bx)$ be as in \eqref{eq:predictive:variance:krig} for Kriging and be $\widehat{RMSE}$ for neural networks and for kernel methods. Then, we consider the classifier, tuned by the parameter $\tau \in \mathbb{R}$, that classifies $\bx$ as unsafe if
\begin{equation} \label{eq:def:lambda}
\hat{f}_{code}(\bx) - \tau \hat{\sigma}_{code}(\bx)
\end{equation}
is smaller than $300 ^ \circ$, and classifies $\bx$ as viable otherwise. One can give a large value to $\tau$, if one considers that classifying as viable an unsafe $\bx$ is more harmful than classifying as unsafe a viable $\bx$, and a small value to $\tau$ otherwise.

\paragraph{Classification Results.}

We present the classification results of the three metamodels in the form of their Receiver Operating Characteristic (ROC) curves (see e.g. \cite[Ch.11.16]{tuffery11data}). For any fixed $\tau$ and for each classifier, we define the ``true unsafe rate'' as the ratio, on the test base, of the number of $\bx$ that are unsafe and classified as unsafe, divided by the number of $\bx$ that are unsafe ($385$). We also define the the ``false unsafe rate'' as the ratio, on the test base, of the number of $\bx$ that are viable and classified as unsafe, divided by the number of $\bx$ that are viable ($1228$). Thus, selecting a sequence of increasing values of $\tau$ yields a sequence of increasing ``false unsafe rate'' values and a sequence of increasing ``true unsafe rate'' values. Plotting the latter sequence as a function of the former constitutes a ROC curve. The higher this curve is, the better the classifier is, since the ``true unsafe rate'' is larger, for a given ``false unsafe rate''.

\begin{figure}
\begin{center}
\includegraphics[angle=0,width=8cm,height=8cm]{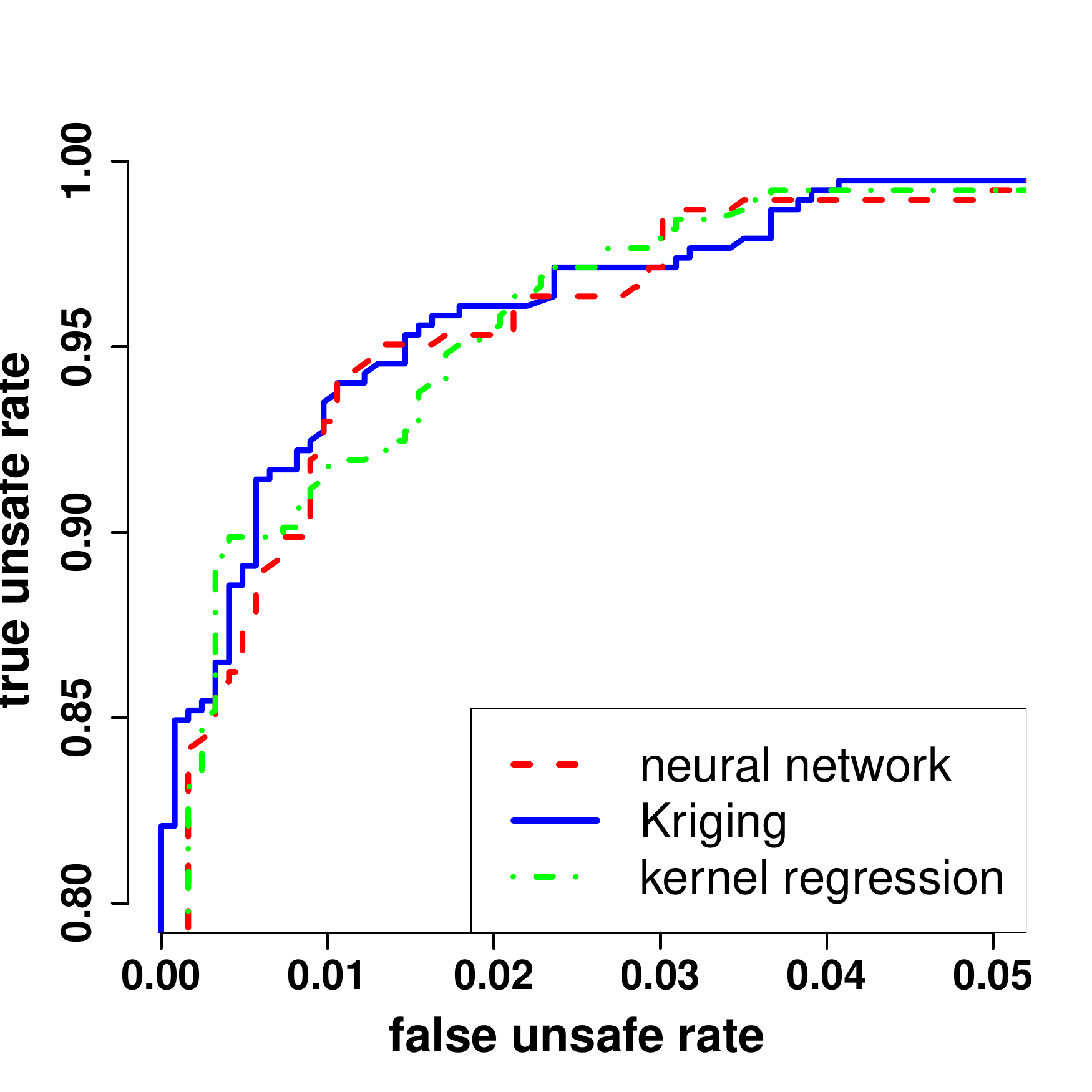}
\end{center}
\caption[ROC curve for the ``Fusion\_Margin'' output]{Plot of the ``true unsafe rate'' as a function of the ``false unsafe rate'', for varying values of the tuning parameter $\tau$ in \eqref{eq:def:lambda}. The total number of unsafe $\bx$ is $385$ and the total number of viable $\bx$ is $1228$. The areas under the ROC curves are $0.9977$ for Kriging, $0.9974$ for the neural networks and $0.9972$ for kernel methods.}
\label{fig:ROC:old:MF}
\end{figure}

The ROC curves for the three metamodels are presented in Figure \ref{fig:ROC:old:MF}. The classification results are good, since one can, for example, achieves more than $95 \%$ ``true unsafe rate'' for less than $5 \%$ ``false unsafe rate''. The three ROC curves are difficult to compare visually, since depending on the value of the ``false unsafe rate'', any of them can be above the others. Nevertheless,
the values of the area under the ROC curves \cite[Ch.11.16]{tuffery11data}, are $0.9977$ for Kriging, $0.9974$ for the neural networks and $0.9972$ for kernel methods, which give the same ranking of the three metamodels, in terms of accuracy, as for the prediction errors.

\section{Improvement of code behaviour} \label{section:outlier:detections}

\subsection{Outlier detection} \label{subsection:outlier:detected:old}

As shown in Figure \ref{fig:pred:old:MF}, for some intputs $\bx_t^{(j)}$ in the test base, the corresponding Germinal output values $f_{code}(\bx_t^{(j)})$ are predicted with particularly large errors. As we show below, similar couples $(\bx^{(j)},f_{code}(\bx^{(j)}))$, that we call outliers, exist in the learning base.

To detect outliers in the learning base, we define the normalized prediction error at $\bx^{(j)}$ as
\begin{equation} \label{eq:normalized_prediction:error}
\frac{  \tilde{f}_{code}(\bx^{(j)}) - f_{code}(\bx^{(j)})  }{ 
\tilde{\sigma}_{code}(\bx^{(j)}) },
\end{equation}
where $\tilde{f}_{code}(\bx)$ is defined for the three metamodels as in \eqref{eq:RMSEl}, and where $\tilde{\sigma}_{code}(\bx^{(j)})$ is defined as $\widehat{RMSE}$ for neural networks and kernel methods and as in \eqref{eq:Kriging:LOO:predictive:variance} for Kriging. For neural networks and kernel methods, the normalization term is $\tilde{\sigma}_{code}(\bx^{(j)}) = \widehat{RMSE}$, so that the average of the squares of \eqref{eq:normalized_prediction:error}, for $j=1,...,n$, is $1$. Under the Gaussian process assumption, the Kriging normalized errors \eqref{eq:normalized_prediction:error} follow the standard Gaussian distribution. [Note that these errors are however not independent in general.]

\begin{figure}
\begin{center}
\begin{tabular}{ccc}
\hspace{-1cm} \includegraphics[angle=0,width=5cm,height=6cm]{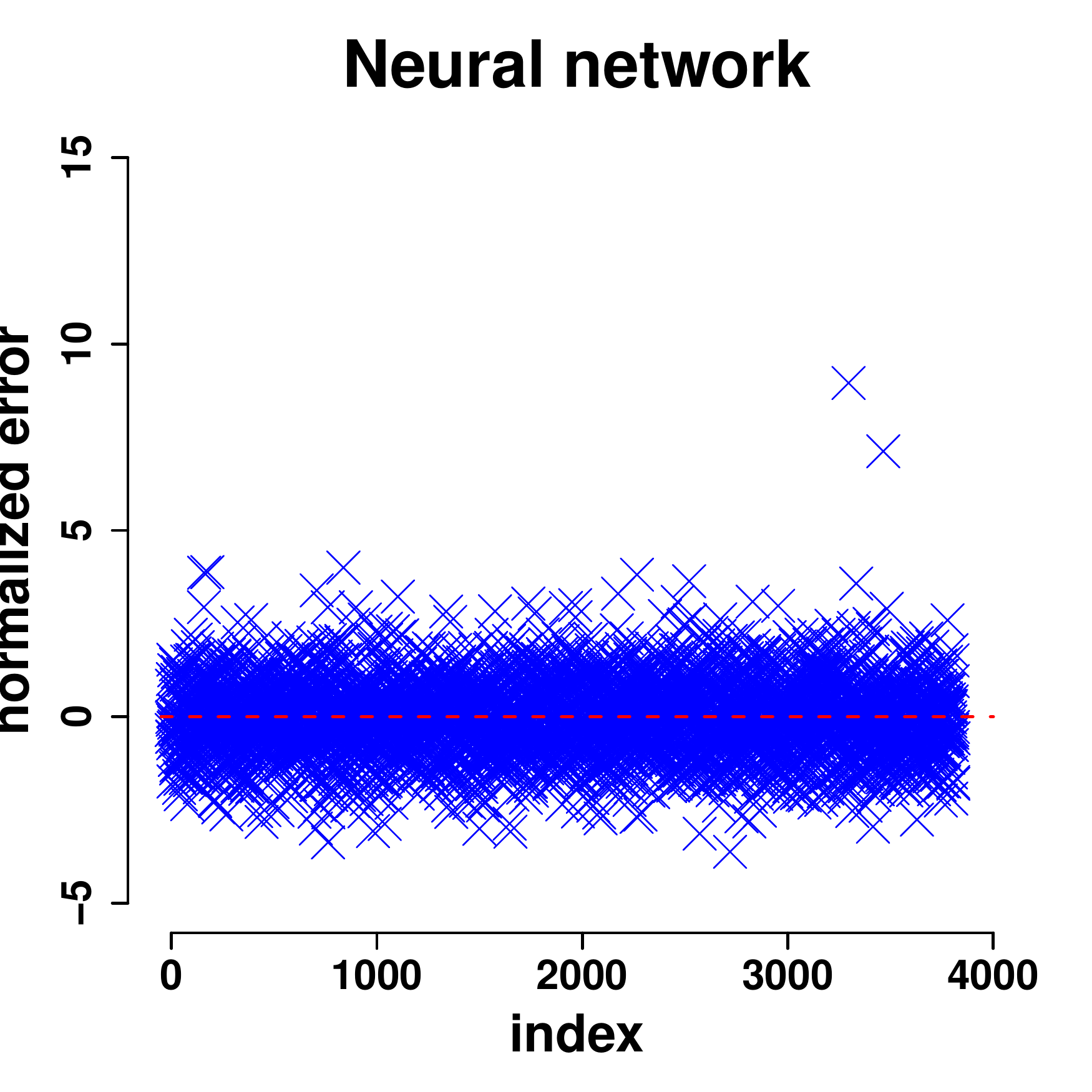}
&
\includegraphics[angle=0,width=5cm,height=6cm]{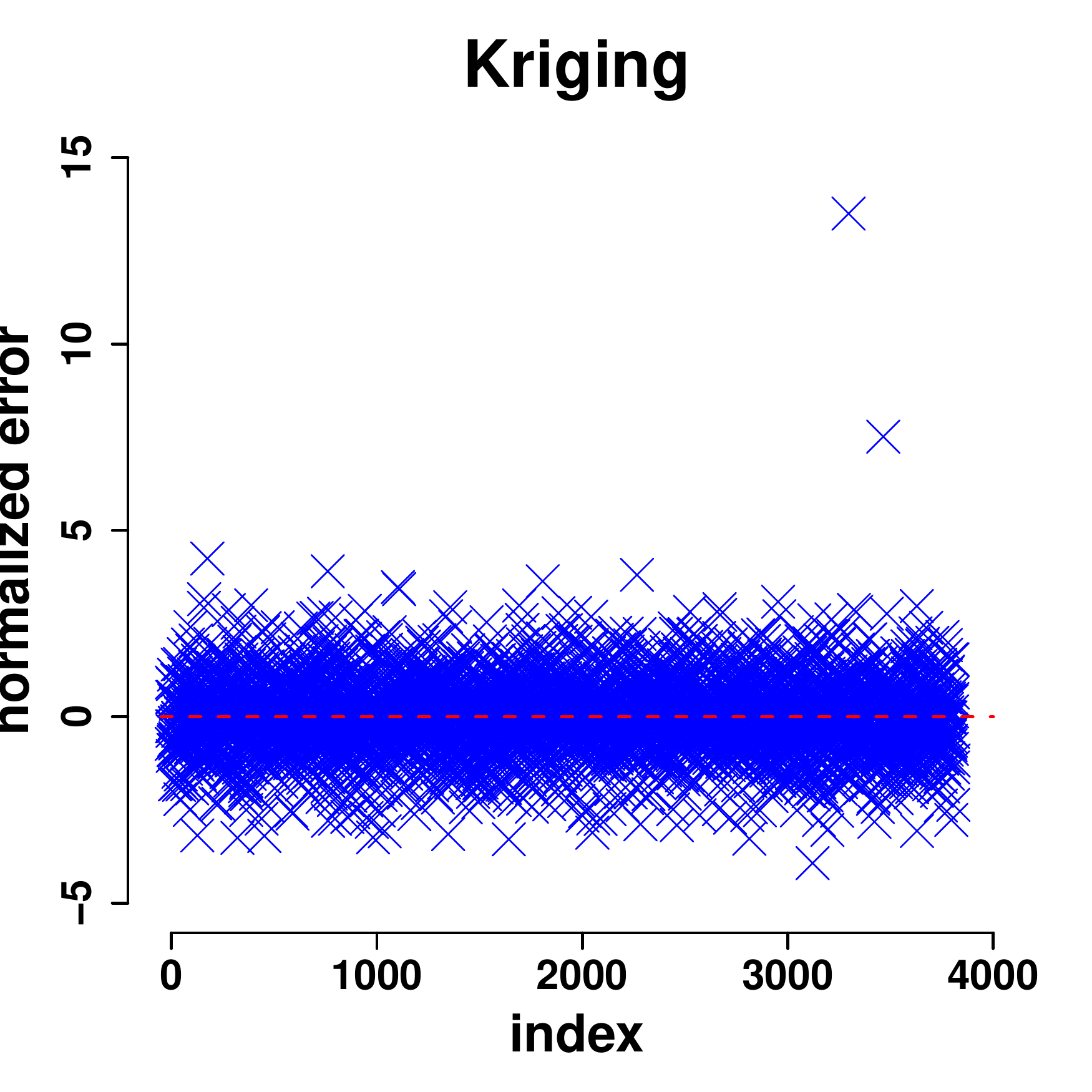}
&
\includegraphics[angle=0,width=5cm,height=6cm]{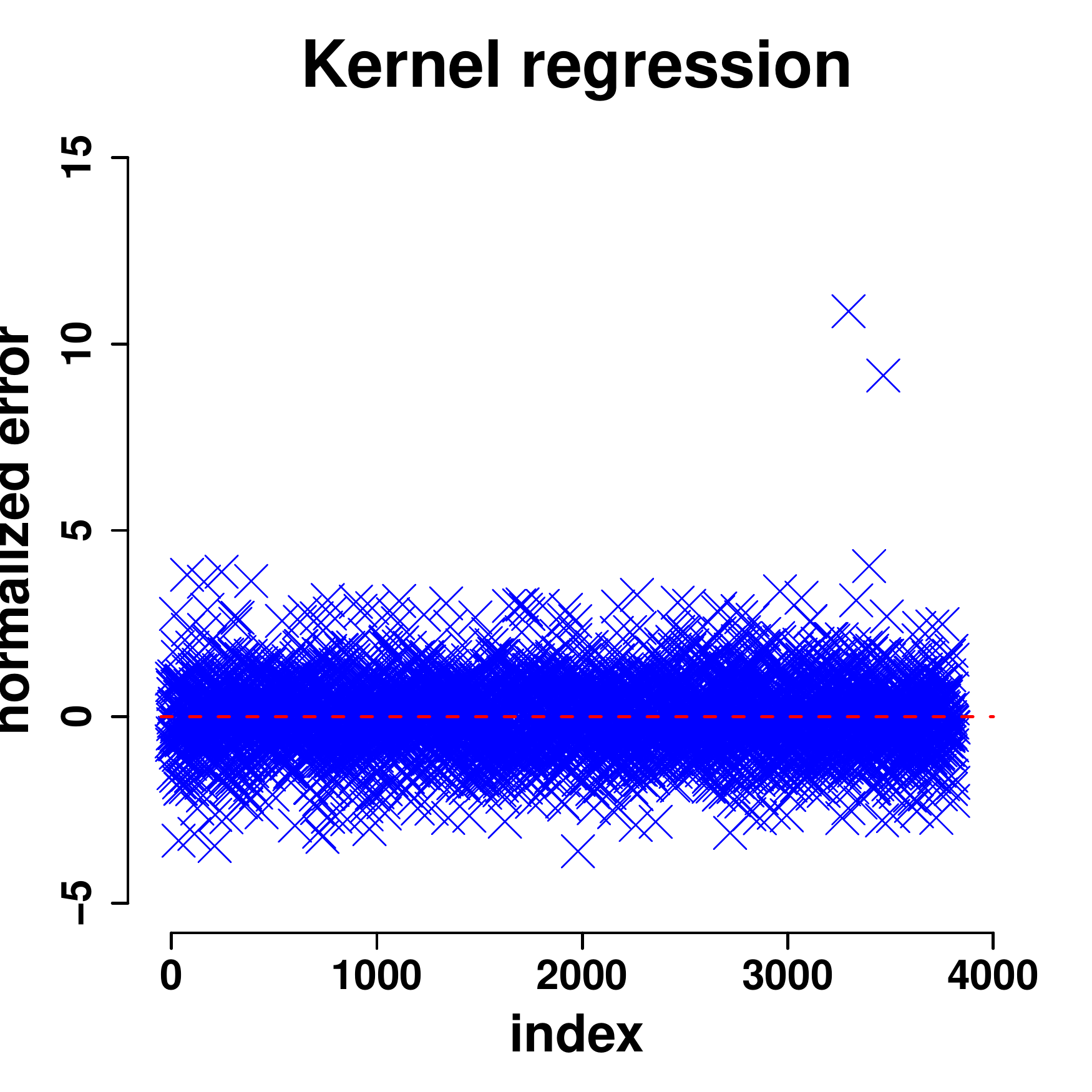}
\end{tabular}
\end{center}
\caption{Plot of the normalized prediction errors \eqref{eq:normalized_prediction:error} on the learning base (y-axis), as a function of the computation index (x-axis), for neural networks (left), Kriging (middle) and kernel methods (right). Case of the original computations. Two outliers stand out and correspond to the same computations for the three metamodels.}
\label{fig:normalized:errors}
\end{figure}

The normalized errors are presented in Figure \ref{fig:normalized:errors}.
For the three metamodels, two particularly large (in absolute value) normalized errors stand out. The other remaining errors are homogeneous and considerably smaller. In addition, for the three metamodels, these two largest errors correspond to the same computations  $(\bx^{(j)},f_{code}(\bx^{(j)}))$. Thus, these two computations are detected as outliers and should be thoroughly studied. The other computations with large normalized errors could possibly benefit from a more detailed investigation, but this is less of a priority.

One should explain why the normalized prediction errors for the two outliers are so large. We think that there are, in general, three possible causes for large prediction errors. The metamodels can be imperfectly specified, the input $\bx^{(j)}$ for the outlier can be isolated in the learning base, and the output $f_{code}(\bx^{(j)})$ can stem from a computation of the Germinal code that have failed, so that the value $f_{code}(\bx^{(j)})$ does not make sense from a physical point of view and is very different from the values $f_{code}(\bx^{(k)})$, for inputs $\bx^{(k)}$ in the learning base that are close to $\bx^{(j)}$.

These three causes can and should be addressed: Metamodels can be questioned and improved (for example the choice of the covariance function in Kriging or the structure of the neural network can be updated). New computations of the code can be made for inputs $\bx$ in areas of the input space $\mathcal{D}$ that are insufficiently covered.
Finally, suspicious computations can be checked manually. For the two aforementioned outlier computations, this third solution is appropriate.

Indeed, a detailed analysis of the two computation output files for the two outliers shows that the same specific warning message was given for both. This message appears in only $14$ of the remaining computations of the learning base. Furthermore, computations with input variables similar to those of the two outliers show drastically different output values. Hence, we can conclude that this warning message implies much more serious consequences on the computation result than indicates its current description (``relatively'' bad numerical behaviour). 

In light of this analysis, we update the code postprocessor by giving a more important weight to this warning message. Computations whose output files contain this message are now labelled as failures and are not incorporated in the learning base. In Section \ref{subsection:reduction:nugget:effect}, we show how the postprocessor can also be updated, and we present the metamodel prediction results for the corresponding improved code manager in Section \ref{section:updated}.

Note that, even though our analysis indicated that the computations have failed numerically for the two outliers, it would be very challenging and time consuming to point out the exact nature of the failure.

\subsection{Reduction of the code instabilities} \label{subsection:reduction:nugget:effect}

\begin{figure}
\begin{center}
 \includegraphics[angle=0,width=14cm,height=6cm]{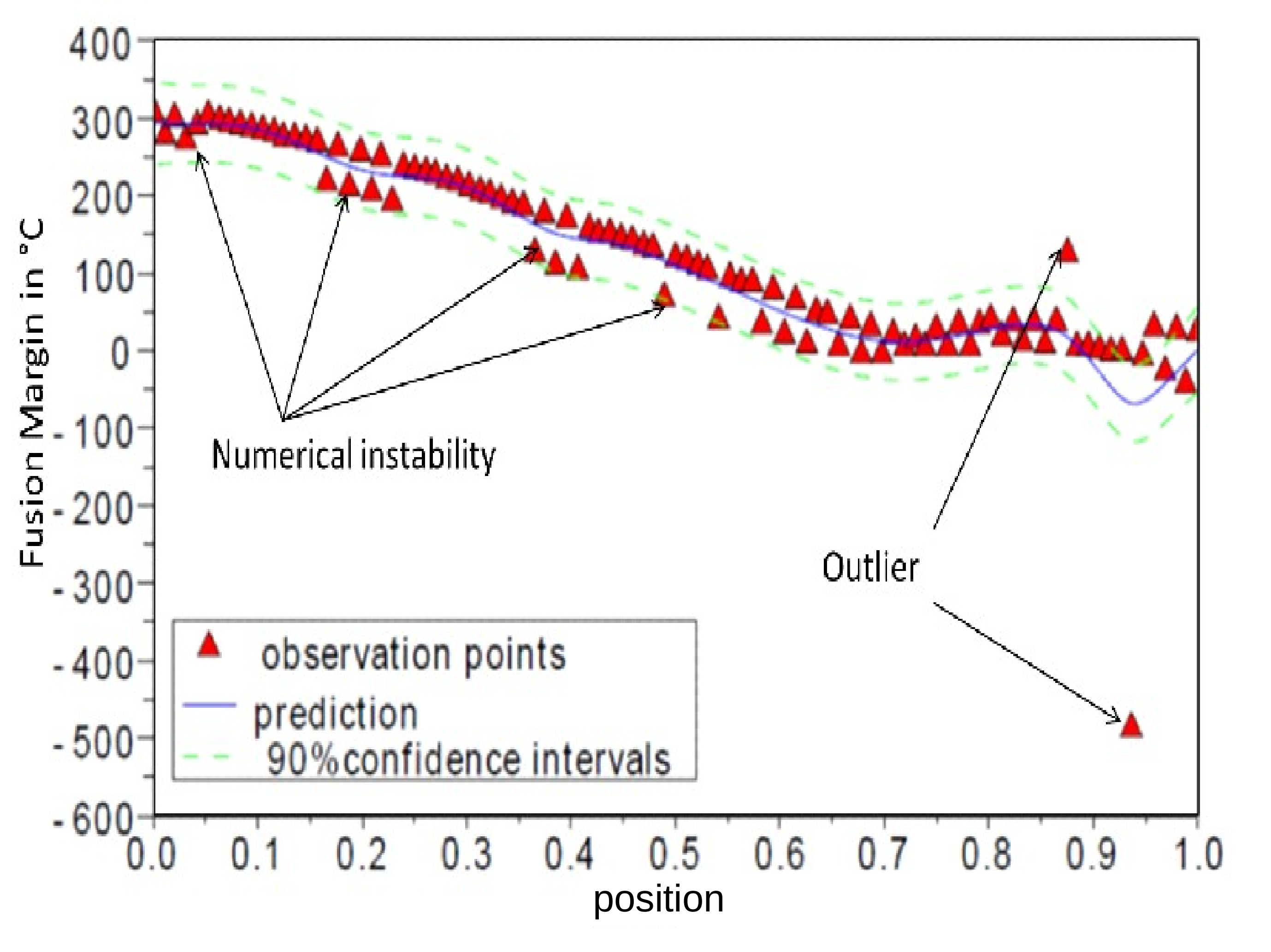}
\end{center}
\caption[]{One-dimensional representation of the Germinal code function. We run $97$ computations, whose input points are located along a line segment, between two points $\boldsymbol{a}$ and $\boldsymbol{b}$ of the (normalized) input space $[0,1]^{11}$. Input points are thus indexed by their position on the segment, where $0$ corresponds to $\boldsymbol{a}$ and $1$ corresponds to $\boldsymbol{b}$. We observe a code instability, causing oscillations of the code response, that have no physical meaning but are caused by a preprocessing issue. We also show the prediction and $95 \%$ confidence intervals obtained with the Kriging estimated covariance parameters of Table \ref{table:Germinal:estimation:MF}, where the support points of \eqref{eq:hatf:kriging} and \eqref{eq:predictive:variance:krig} correspond to these $97$ computations. The output files of the two outlier computations contain the same warning as that described in Section \ref{subsection:outlier:detected:old}.}
\label{fig:one:dim:function}
\end{figure}

As discussed in Section \ref{subsection:prediction:results}, the relatively large nugget effect estimated by Kriging ($(28.5 ^\circ)^2$) is a sign of code instabilities. [Note, in comparison, that the  designed numerical approximations of Germinal, including for instance rounding of values, are around $1^\circ$.] To obtain graphical information on these instabilities, we run $97$ additional computations, whose input points are located along a line segment of the (normalized) input space $[0,1]^{11}$, and can hence be ordered. This provides us a one-dimensional visualization of the Germinal code function that we show in Figure \ref{fig:one:dim:function}.

In Figure \ref{fig:one:dim:function}, we observe oscillations of the code response, that can clearly not correspond to the modelled physical process (typically assumed to entail piecewise differentiable functions), and are hence code instabilities. [Note that we also observe two outlier computations, whose output files actually contain the same warning as that described in Section \ref{subsection:outlier:detected:old}.] For the sake of illustration, we also show the prediction and $95 \%$ confidence intervals obtained with the Kriging estimated covariance parameters of Table \ref{table:Germinal:estimation:MF}, where the support points (used to construct $\br(\bx)$, $\R$ and $\by$) of \eqref{eq:hatf:kriging} and \eqref{eq:predictive:variance:krig} correspond to the aforementioned $97$ additional computations. We see that the covariance parameters estimated by Kriging, and in particular the nugget effect, are appropriate and adapted to the code instabilities, and entail satisfactory prediction and confidence intervals.

\begin{figure}
\begin{center}
 \includegraphics[angle=0,width=14cm,height=6cm]{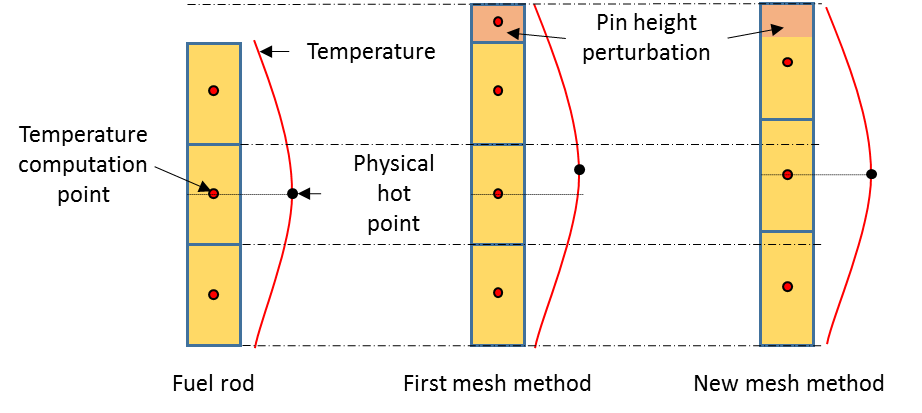}
\end{center}
\caption[]{Simplified illustration of the impact of a perturbation of the pin height input on the mesh generation for a Germinal computation. In the original Germinal computations, mild changes of the pin height can cause significant modifications of the mesh, themselves causing the code instabilities of Figure \ref{fig:one:dim:function}. We consequently updated the preprocessor to solve this issue.}
\label{fig:mesh:shift}
\end{figure}

We investigated closely consecutive computations in the code instability zones of Figure \ref{fig:one:dim:function}.
We found out that the code preprocessor generates automatically an axial mesh from a global pin height. A small variation of the pin height changes the fusion margin only moderately but may change the location of the maximum-temperature space point (the physical hot point) much more significantly. As illustrated in Figure \ref{fig:mesh:shift}, with the current mesh method, a computation point (a mesh node) coinciding with the physical hot point can shift away from it with a small pin height variation, thus yielding a computation of the fusion margin that is numerically (and erroneously) overly different.

\begin{figure}
\begin{center}
 \includegraphics[angle=0,width=14cm,height=6cm]{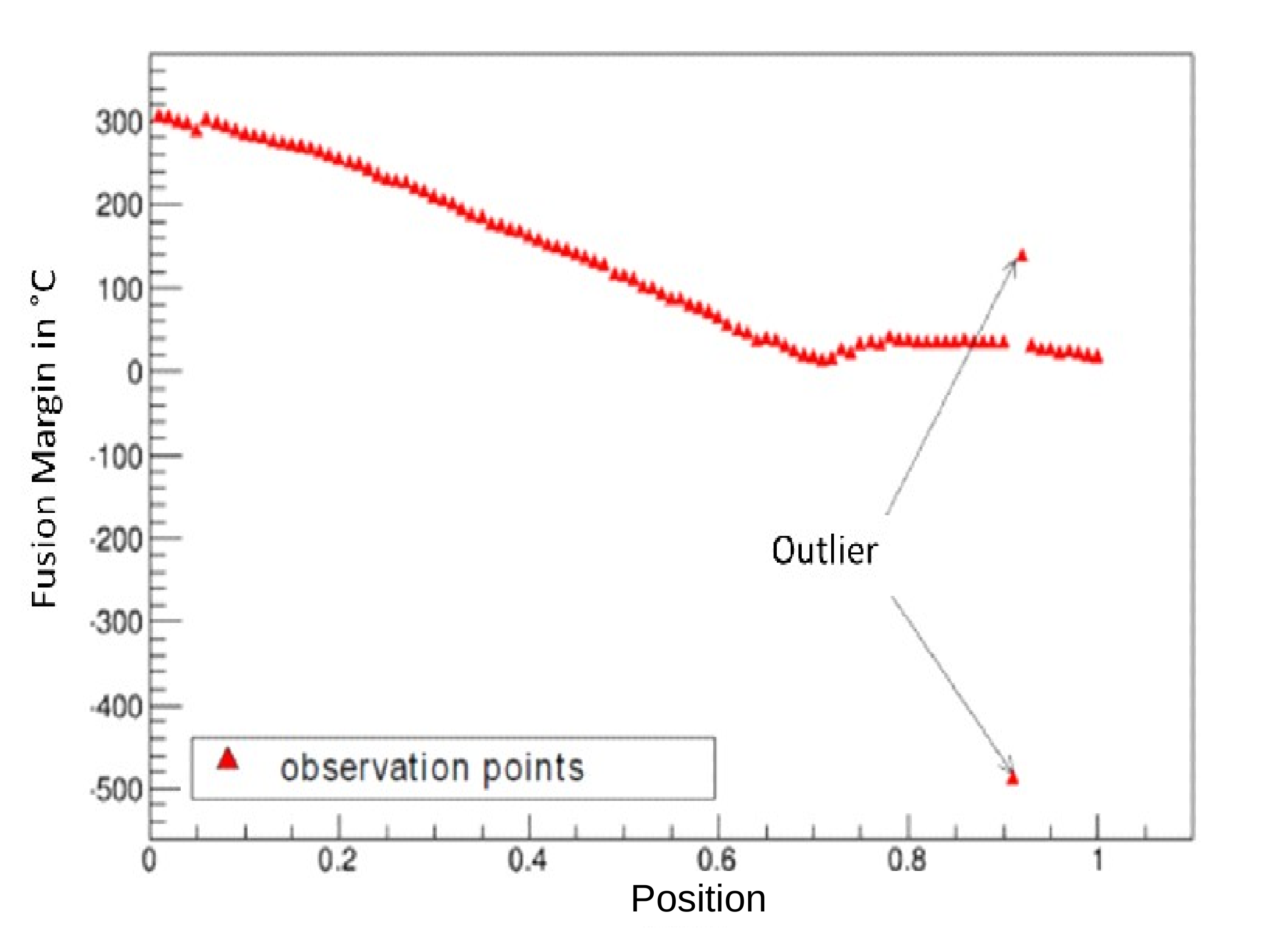}
\end{center}
\caption[]{Same settings as in Figure \ref{fig:one:dim:function}, but for the updated Germinal computations. The code instabilities have been corrected.}
\label{fig:one:dim:function:deux}
\end{figure}

Consequently, we updated the preprocessor, as illustrated in Figure \ref{fig:mesh:shift}. Together with the previously discussed postprocessor update, this yields an updated version of the Germinal code manager, that we used to generate new output values for the inputs of Figure \ref{fig:one:dim:function}. Figure \ref{fig:one:dim:function:deux} shows that the code instabilities have been corrected in the new code version. Hence, eventually, the Kriging nugget effect helps detecting code instabilities that can then be investigated and corrected.

\section{Prediction and classification results for the updated Germinal computations} \label{section:updated}

\subsection{Prediction results}

A discussed in Section \ref{section:outlier:detections}, the Germinal code manager has been updated. We have repeated all the Germinal computations for the input points of the original learning and test bases of Section \ref{section:original}. These inputs, together with the new output values, correspond to updated learning and test bases that we use in this section. Because the postprocessor has been updated, additional computations are flagged as failure (those presenting the warning message discussed in Section \ref{subsection:outlier:detected:old}), so that the updated learning and test bases have $3791$ and $1606$ points. The metamodels are then used exactly as for the original learning and test bases.

\paragraph{Estimated covariance parameters for Kriging.}

For the updated learning base, the estimated correlation lengths of the Kriging metamodel are similar to those for the original learning base. However, the estimate of the nugget variance $\hat{\delta}^2$ decreases significantly from the original to the updated learning base, going from $(28.5 ^{\circ} )^2$ (Table \ref{table:Germinal:estimation:MF}) to $(19.8 ^{\circ} )^2$ . This is a sign that the pre- and post-treatment procedures for the Germinal code have been improved, as is illustrated by Figure \ref{fig:one:dim:function:deux}. Nevertheless, since the nugget variance remains significant, we believe that pre- and post-treatment issues might remain. The prediction results presented below are in agreement with this discussion.

\paragraph{Prediction results.}

The prediction results, for the updated learning and test bases, are given in Table \ref{table:prediction:results:MF:new}.
The standard deviation of the output on the test base is $326 ^{\circ}$, and the RMSE for neural networks, Kriging and kernel methods are respectively $31.3 ^{\circ}$, $27.6 ^{\circ}$ and $38.5 ^{\circ}$. Hence, the prediction errors of the metamodels are smaller than for the original computations in Section \ref{section:original}, but still are of comparable order of magnitude. 
This observation, together with the updated estimate of the nugget variance, indicates that the code instabilities have been reduced but not suppressed. 

As for the original computations, Kriging gives the smallest RMSE, followed by neural networks and kernel methods. The quantity $\widehat{RMSE}$ is almost a perfect estimator of RMSE for Kriging and kernel methods and is again slightly too optimistic for neural networks. The $90\%$ confidence intervals provided by Kriging are also appropriate, as they contain $91.2\%$ of the output values in the test base (CIR = $91.2\%$ in \eqref{eq:Germinal:CIR}).

\begin{table}
\begin{center} 
\begin{tabular}{|c | c  |  c  | c | c | c | c |}
\hline 
 & $\widehat{RMSE}$ & RMSE & $\hat{Q}^2$ & $Q^2$ & $q_{0.9}$ & $q_{0.95}$  \\
\hline
Neural network &  $27.5 ^{\circ}$  & $31.3 ^{\circ}$ & $0.993$ & $0.991$ & $48.7 ^{\circ}$  & $63.4 ^{\circ}$  \\ 
\hline
Kriging &  $27.2^{\circ}$ &  $27.6 ^{\circ}$  & $0.993$ & $0.993$ & $43.2 ^{\circ}$  & $54.0 ^{\circ}$  \\
\hline
Kernel methods  & $38.3 ^{\circ}$ &  $38.5 ^{\circ}$ & $0.986$ & $0.986$ & $60.8 ^{\circ}$ & $75.3 ^{\circ}$  \\
\hline
\end{tabular}
\end{center} 
\caption{Same context as for Table \ref{table:prediction:results:MF} but for the updated Germinal computations. The standard deviation of the output on the test base is $326.2 ^{\circ}$. The estimates $\widehat{RMSE}$ and $\hat{Q}^2$ are more accurate for Kriging and kernel methods than for the neural networks, thanks to the virtual LOO formulas. 
}
\label{table:prediction:results:MF:new} 
\end{table}

\subsection{Classification}

\begin{figure}
\begin{center}
\begin{tabular}{cc}
\includegraphics[angle=0,width=7cm,height=7cm]{OLD_MF_ROC_standard_ter.pdf}
&
\includegraphics[angle=0,width=7cm,height=7cm]{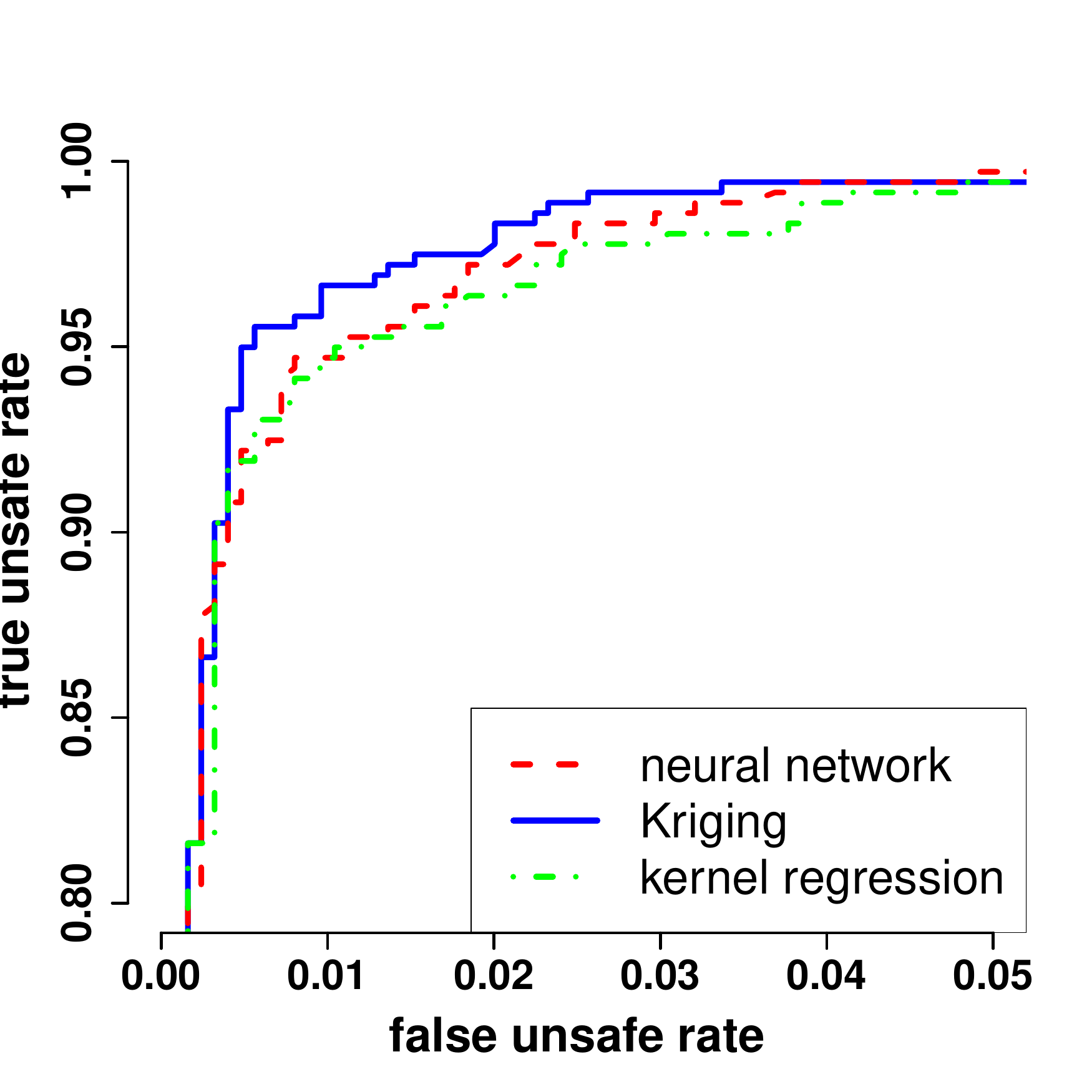}
\end{tabular}
\end{center}
\caption[ROC curve for the ``Fusion\_Margin'' output]{Plot of the ``true unsafe rate'' as a function of the ``false unsafe rate'', for varying values of the tuning parameter $\tau$ in \eqref{eq:def:lambda}, for the original computations (left) and the updated computations (right).
The total number of unsafe $\bx$ is $385$ (left) and $359$ (right), and the total number of viable $\bx$ is $1228$ (left) and $1247$ (right). For the updated computations, Kriging becomes more accurate than neural networks and kernel methods.}
\label{fig:ROC:new:MF}
\end{figure}

The ROC curves for the three metamodels for the updated computations are presented in Figure \ref{fig:ROC:new:MF}, where we also re-plot the ROC curve of Figure \ref{fig:ROC:old:MF} (original computations) for comparison. In line with the prediction improvement in Table \ref{table:prediction:results:MF:new}, the ROC curves are higher for the updated computations, which indicates that the classifiers perform better. Similarly, the area under the ROC curves are now $0.9984$ for Kriging, $0.9980$ for neural networks and $0.9978$ for kernel methods. Hence the three classifiers have improved performances compared to the original computations.
For the updated computations, the ROC curve of Kriging is more clearly above the ROC curves of neural networks and kernel methods.

Note finally that, from Figure \ref{fig:ROC:new:MF} and Table \ref{table:prediction:results:MF:new}, Kriging performs better, in comparison to neural networks and kernel methods, for the updated computations than for the original ones. Indeed, first, the ratios of the RMSE of Kriging divided by the RMSE of neural networks and kernel methods are smaller in Table \ref{table:prediction:results:MF:new} than in Table \ref{table:prediction:results:MF}. Second, the Kriging ROC curve becomes clearly higher than the two other ones for the updated computations. Similarly, the ratio of the RMSE of neural networks divided by the RMSE of kernel methods is smaller in Table \ref{table:prediction:results:MF:new}  than in Table \ref{table:prediction:results:MF} .

Hence, the relative differences between the three metamodel prediction errors are more accentuated for the new computations than for the old computations. We believe that this holds because of the decrease of the code instabilities in the new computations. Indeed, intuitively, the code instabilities cause systematic prediction errors, stemming from the fact that the brusque changes of $f_{code}(\bx)$ for very small changes of $\bx$ are not predictable. [The square of these prediction errors have values $\hat{\delta}^2$ on average under the Gaussian process model, see \eqref{eq:interpretation:nugget:prediction}.] These systematic errors are the same for the three metamodels, so that, when they become large, the ratios of prediction errors between different metamodels become closer to one.

\section{Conclusion}

Many studies in nuclear engineering, such as optimal conception, require an extensive use of computer codes, for many different input conditions. In order to limit the computation time, computer codes are replaced by metamodels, that provide approximations of the code output values, for a much cheaper computational cost.

In this paper, we present a detailed case study of the metamodeling of the fusion margin output of the Germinal code, in the case of the thermo-mechanical simulation of a fuel pin under irradiation. We compare the metamodels obtained from neural networks, Kriging and kernel methods. In our study, the computation time for metamodel evaluation is similar for Kriging and kernel methods and is the smallest for neural networks. The most accurate predictions are obtained from Kriging, followed by those obtained from neural networks, and finally by those obtained from kernel methods. Kriging and kernel methods provide the most reliable estimates of their prediction errors. This is thanks to the Leave-One-Out formula, which are not directly available for the neural networks. Kriging also arguably provides the most interpretability, with the underlying Gaussian process model and the covariance parameters. The kernel methods are the simplest to implement, and the fit of the corresponding metamodel is the fastest.

Beyond this comparison, we demonstrate the pertinence of these three metamodeling techniques to improve the behaviour of the Germinal code in a design of experiments. Indeed, as many simulation codes, the Germinal code is conceived to be used for a limited number of specific situations, in which experts in physics or numerical simulation dedicate a consequent time to specify the simulation conditions and to interpret the results. In a design of experiments, where here thousands of simulations are carried out, an automatic code manager, consisting in pre- and post-processing scripts, has to replace this human intervention. Hence, specific problems and errors arise in the use of this code manager, that metamodeling techniques can detect, quantify and contribute to correct. 

In the case study we address, we distinguish two types of issues related to the use of a code manager. First, some of the simulations in the design of experiments can be plagued by numerical flaws, that are not flagged by the code manager and which cause the simulation outputs to be meaningless. These meaningless simulation results are well-detected by the metamodels: In our study, the three metamodels detect the same two simulations as doubtful, and a human intervention indeed confirm that computational failures occurred. This property of the metamodels to rank the simulations according to statistical estimates of their reliability is very attractive. Indeed, it is not possible to check manually all the simulations that are carried out, but it is possible to do so for a few simulations that are automatically detected.

The second issue related to the use of a code manager is the instability of the preprocessing step. In our study, we have analysed that modifying input conditions very slightly can cause a non-negligible change in the preprocessing step (e.g. a significantly different mesh), this change then causing a significant variation in the simulation result. This code instability is problematic because it increases the prediction errors of the metamodels. We find that the estimate of the nugget variance provided by Kriging is particularly efficient for detecting and quantifying code instability. Especially, this nugget variance estimate decreases between the original and updated Germinal computations, which coincides with an improvement of the preprocessing step, this improvement then enabling more accurate predictions for the three metamodels. Nevertheless, the nugget variance remains non-negligible for the updated Germinal computations, which is a signal that the code manager can still be improved. 

Once the global presence of code instabilities is detected and quantified, we consider as a rather open problem the question of using metamodels to help code experts to solve them. In this study, we have proposed to carry out computations in a segment of the input space (see Figure \ref{fig:one:dim:function}), in order to have visual information on the code instabilities. This method enables us to detect pairs of very close input conditions yielding significantly different simulation results and to investigate them in details. It would be interesting to see if metamodels can provide more automatic tools to isolate such pairs of input conditions automatically.

Finally, we believe that the above-described issues, arising from the use of the Germinal code in a design of experiments, also occur in a large variety of situations in numerical simulation, in which codes are used automatically for a large number of different simulation conditions.

\section*{Acknowledgements}

The authors would like to thank Guillaume Damblin, Chunyang Li, Cyril Patricot and Am\'elie Rouchon for valuable comments and suggestions.

\FloatBarrier

\newpage

\bibliographystyle{unsrt}
\bibliography{Biblio} 

\end{document}